\documentclass[useAMS]{cJAS2e}

\pdfoutput=1	
\usepackage{ifpdf}
\ifpdf
  \DeclareGraphicsExtensions{.pdf,.jpeg,.png}
\else
  \DeclareGraphicsExtensions{.eps}
\fi
\graphicspath{{./}{./figs/}}
\usepackage{multirow}	
\usepackage{lineno}		

\RequirePackage[colorlinks,citecolor=blue,urlcolor=blue]{hyperref}	
\newcommand{\beq}{%
\ifLineNumbers \begin{linenomath}\begin{equation}
\else \begin{equation} \fi}
\newcommand{\eeq}{%
\ifLineNumbers \end{equation}\end{linenomath}
\else \end{equation} \fi}
\newcommand{\bea}{%
\ifLineNumbers \begin{linenomath}\begin{eqnarray}
\else \begin{eqnarray} \fi}
\newcommand{\eea}{%
\ifLineNumbers \end{eqnarray}\end{linenomath}
\else \end{eqnarray} \fi}
\newcommand{\bes}{%
\ifLineNumbers \begin{linenomath}\begin{subequations}\begin{eqnarray}
\else \begin{subequations}\begin{eqnarray} \fi}
\newcommand{\ees}{%
\ifLineNumbers \end{eqnarray}\end{subequations}\end{linenomath}
\else \end{eqnarray}\end{subequations} \fi}
\newcommand{\mbf}[1]{\mathbf{#1}}
\newcommand{\mrm}[1]{\mathrm{#1}}
\newcommand{\msf}[1]{\mathsf{#1}}

\newcommand{\del}{\nabla}

\newcommand{\dsub}[1]{\partial_{#1}}

\newcommand{\given}{ \, \vert \,}

\newcommand{\p}[2]{p^{#1}_{#2}}
\newcommand{\q}[2]{q^{#1}_{#2}}

\begin{document}

\doi{10.1080/02664763.YYYY.XXXXXX}
\issn{1360-0532}
\issnp{0266-4763}
\jvol{00} \jnum{00} \jyear{20XX} \jmonth{XXXXXX}

\markboth{Taylor \& Francis and I.T. Consultant}{Journal of Applied Statistics}

\articletype{ARTICLE}

\title{Prediction of particle type from measurements of particle location: A physicist's approach to Bayesian classification}

\author{Robert W. Johnson$^{\ast}$ \thanks{$^\ast$Corresponding author. Email: robjohnson@alphawaveresearch.com\vspace{6pt}}\\
\vspace{6pt} {\em{Alphawave Research, Jonesboro, GA 30238}}
\vspace{6pt}\received{\today} }

\maketitle

\begin{abstract}
The Bayesian approach to the prediction of particle type given measurements of particle location is explored, using a parametric model whose prior is based on the transformation group.  Two types of particle are considered, and locations are expressed in terms of a single spatial coordinate.  Several cases corresponding to different states of prior knowledge are evaluated, including the effect of measurement uncertainty.  Comparisons are made to nearest neighbor classification and kernel density estimation.  How one can evaluate the reliability of the prediction solely from the available data is discussed.
\bigskip
\begin{keywords}
Bayesian classification; Inductive reasoning; Machine learning\\
{\bf MSC}:  62F15; 62H30; 62M30
\end{keywords}\bigskip
\end{abstract}


\nolinenumbers

\section{Introduction}
A common problem that appears in many guises is how to predict the type of some unidentified particle knowing only its location, given a list of locations for where such particles of identified type have been found.  One example of such a problem is the prediction of whether a passerby crossing some particular line in space is male or female based upon knowledge of where on the line previous passersby of known gender have crossed.  The problem is predicated on the assumption that measurements of location are inexpensive compared to measurements of classification, so that after an initial training set of data has been evaluated some algorithm can then be used to predict the classification based only upon a given location with some degree of certainty that can be determined.

A specific case of this problem has recently been addressed using the method of $k$th nearest neighbor classification by Hall, \textit{et al.}~\cite{hall-5276H}.  In that paper, an optimal choice of $k$ is evaluated empirically from the data, allowing for a nonparametric estimate of the desired probability.  Another approach commonly employed for problems of this type is kernel density estimation~\cite{eberts-2013,kim-2529,terrell-1236}.  In this paper, we will examine how a parametric model is integrated over the parameter manifold with respect to the evidence measure to yield a prediction which depends only upon the given location.  The transformation group aspect of the algorithm refers to the use of uninformative priors for the parameters, which nonetheless may be non-uniform for some particular choice of coordinate mapping of the parameter manifold.  An analysis similar in spirit to this one has been presented by Poitevineau, \textit{et al.}~\cite{Poitevineau-2010724}.

The process of inductive reasoning is best described using the language of conditional probability theory~\cite{Bretthorst-1988,Durrett-1994,Sivia-1996}.  Let us quickly review the notation and nomenclature that will be used in this paper.  The formal statement of the expression for the probability of $A$ given conditions $B$ can be written as \beq
p ( A \given B) \equiv \p{A}{B} \;,
\eeq where $A$ and $B$ can have arbitrary dimensionality; for example, $A$ could be a vector of measurements, and $B$ could include both the vector of parameters associated with some model as well as any other conditioning statements such as the model index.  The sum and product rules of probability theory yield the expressions for marginalization and Bayes' theorem, \bea
\p{A}{} &=& \int_{\{B\}} \p{A, B}{} \, dB \;, \\
\p{B}{A} \p{A}{} &=& \p{A}{B} \, \p{B}{} \;,
\eea where marginalization follows from the requirement of unit normalization, and Bayes' theorem follows from requiring logical consistency of the joint density $\p{A, B}{} = \p{B, A}{}$.  Certain names have come to be associated with the various factors above, but as Sivia~\cite{Sivia-1996} points out, what one calls a probability is irrelevant, as the distinction between $\p{A}{B}$ and $\p{A}{C}$ is carried explicitly by the differences in the conditioning statements.  Nonetheless, the terms ``likelihood'' and ``prior'' are useful for describing how new data updates one's state of knowledge.  Instead of the term ``posterior'' for the estimate of the parameter probability we will use ``evidence'', and the chance of measuring the data based on no other knowledge will not be named as it is not necessary for the normalization of the evidence measure nor for the evaluation of the relative evidence for competing models.

\section{Definition of the model}
For consistency of comparison, we will follow as closely as possible the notation used by Hall, \textit{et al.}~\cite{hall-5276H}.  The population of particles decomposes into two classifications, denoted type $X$ and type $Y$, and the location for each particle type is assumed to follow an independent normal distribution in the spatial dimension $z$.  The measured locations of the particles identified as type $X$ can be expressed as the vector $\mbf{X} \equiv X_j$ for integer $j \in [1, J]$, and similarly $\mbf{Y} \equiv Y_k$ for $k \in [1, K]$, such that $N = J + K$ gives the total number of classified particles.  The location measurements for type $X$ are assumed to be drawn from the normal distribution $\p{X_j}{\bar{f}, \tilde{f}} \equiv f(X_j)$, where \beq
f(z) = ( 2 \pi \tilde{f}^2 )^{-1/2} \exp^{-1/2} [ ( \bar{f} - z ) / \tilde{f} ]^2 \; ,
\eeq and similarly for $\p{Y_k}{\bar{g}, \tilde{g}} \equiv g(Y_k)$, using notation $\exp^\alpha (\beta) \equiv e^{\alpha \beta}$ with an economy of brackets when possible.  One may alternately interpret that equation to mean that the standard deviation of the location measurement process is equal to $\tilde{f}$ (or $\tilde{g}$).  The relative likelihood of a particle being a $Y$ rather than an $X$ is denoted by the parameter $m$, such that the absolute likelihood of being an $X$ is $\p{X}{m} = (1+m)^{-1}$.  The parametric model, then, for the probability that some new, unclassified datum is of type $X$ knowing only its location $z$ is \beq
\p{X \mathrm{\,at\,} z}{z, m, \bar{f}, \bar{g}, \tilde{f}, \tilde{g}} = \left[ 1 + m\, \zeta(z) \right]^{-1} \; ,
\eeq where $\zeta(z) \equiv g(z) / f(z)$, and by normalization $\p{Y \mathrm{\,at\,} z}{z, m, \bar{f}, \bar{g}, \tilde{f}, \tilde{g}} = 1 - \p{X \mathrm{\,at\,} z}{z, m, \bar{f}, \bar{g}, \tilde{f}, \tilde{g}}$.  The desired quantity, however, is the estimate of that likelihood given the data $\p{X \mathrm{\,at\,} z}{z, N, J, K, \mbf{X}, \mbf{Y}}$.  Inductive reasoning is used to relate these quantities of interest.

For brevity of notation, any knowledge that can be derived from the conditioning statements explicitly present will be suppressed, \textit{e.g.} $\p{X \mathrm{\,at\,} z}{z, N, J, K, \mbf{X}, \mbf{Y}} = \p{X \mathrm{\,at\,} z}{z, \mbf{X}, \mbf{Y}}$.  Let us also collect the coordinates of the parameter manifold into the contravariant position vector $\mbf{r} \equiv ( m, \bar{f}, \bar{g}, \tilde{f}, \tilde{g} )$, such that $\del \equiv \partial / \partial \mbf{r}$ is a covariant vector.  By marginalization, the desired quantity can be written as the expectation value of the observable as a function of the parameters weighted by the evidence measure integrated over the entire parameter manifold, not simply taken from its most likely value at the mode, formally expressed as \bes
\p{X \mathrm{\,at\,} z}{z, \mbf{X}, \mbf{Y}} &=& \int_{\{\mbf{r}\}} \p{X \mathrm{\,at\,} z, \mbf{r}}{z, \mbf{X}, \mbf{Y}} d\mbf{r} \\
 &=& \int_{\{\mbf{r}\}} \p{X \mathrm{\,at\,} z}{z, \mbf{r}} \p{\mbf{r}}{\mbf{X}, \mbf{Y}} d\mbf{r} \equiv \langle \p{X \mathrm{\,at\,} z}{z, \mbf{r}} \rangle_{\mbf{r} \given \mbf{X}, \mbf{Y}} \; ,
\ees where the conditioning on $N$ is implicit.

The evidence measure decomposes into a product of factors according to what knowledge is required for their determination $\p{\mbf{r}}{\mbf{X}, \mbf{Y}} \propto \p{m}{J, K} \p{\bar{f}, \tilde{f}}{\mbf{X}} \p{\bar{g}, \tilde{g}}{\mbf{Y}}$, which themselves factor into products of likelihoods and priors.  For example, the evidence density for $m$ is $\p{m}{J, K} \propto \p{J, K}{m} \p{m}{}$, and similarly for the remaining factors.  Addressing first the likelihood factors, the chance of observing $K$ out of $N$ particles of type $Y$ given $m$ is \beq
\p{J, K}{m} = m^K / (1+m)^{J+K} = m^K (1+m)^{-N} \; ,
\eeq whereas the chance of observing locations $\mbf{X}$ given values for $\bar{f}$ and $\tilde{f}$ is \beq
\p{\mbf{X}}{\bar{f}, \tilde{f}} = \prod_{z \in \mbf{X}} f(z) = (2 \pi \tilde{f}^2)^{-J/2} \exp^{-1/2} \sum_{z \in \mbf{X}} [ ( \bar{f} - z) / \tilde{f} ]^2 \; ,
\eeq and similarly $\p{\mbf{Y}}{\bar{g}, \tilde{g}} = \prod_{z \in \mbf{Y}} g(z)$.  The likelihood of the data at manifold position $\mbf{r}$ is then given by their product $\p{\mbf{X}, \mbf{Y}}{\mbf{r}} = \p{J, K}{m} \p{\mbf{X}}{\bar{f}, \tilde{f}} \p{\mbf{Y}}{\bar{g}, \tilde{g}}$.

Next let us look at the prior factors $\p{\mbf{r}}{} = \p{m}{} \p{\bar{f}, \tilde{f}}{} \p{\bar{g}, \tilde{g}}{}$.  The transformation group approach to selecting an uninformative prior is based on the principle of indifference as represented by the requirement of consistency under various transformations of the parameter or data coordinate mappings.  Dose~\cite{dose-350} gives an excellent description of the process.  According to Jaynes~\cite{jaynes-1968}, the prior suggested by Jeffreys for the parameters of a Gaussian distribution results from satisfaction of the functional equation for transformations in location and scale, thus $\p{\bar{f}, \tilde{f}, \bar{g}, \tilde{g}}{} \propto (\tilde{f} \tilde{g})^{-1}$.  In that paper, he argues that the same functional form is appropriate for the rate parameter of a Poisson process, which will be called $n$.  In the presentation by Hall, \textit{et al.}~\cite{hall-5276H}, the arrival of the particles at the axis of measurement is assumed to follow the Poisson distribution with parameters $\mu$ and $\nu$ for types $X$ and $Y$ respectively, thus the total number of particles expected in one unit of time is $n = \mu + \nu$.  Writing $m = \nu / \mu$, the determinant of the Jacobian is $\det \msf{J} = n / (1+m)^2$, leading to the transformation of the prior $\p{n, m}{} = \p{\mu, \nu}{} \det \msf{J}$, whereby \beq
\p{m}{} \propto n^2 (1+m)^{-2} \mu^{-1} \nu^{-1} = m^{-1} \; ,
\eeq which states that $m$ is just as likely to be between 0.1 and 1 as it is to be between 1 and 10 before any observations are recorded.  The prior measure for the parameter manifold is thus $\p{\mbf{r}}{} \propto (m \tilde{f} \tilde{g})^{-1}$, where the constant of proportionality is determined by the limits of consideration.  In particular, finite limits for $m$ must be symmetric in scale about unity so that the prior expectation of finding a particle of type $X$, \beq
\langle \p{X}{m} \rangle_{m \given m_\infty} \equiv \dfrac{\int_{1/m_\infty}^{m_\infty} m^{-1} (1+m)^{-1} dm}{\int_{1/m_\infty}^{m_\infty} m^{-1} dm} \; ,
\eeq remains equal to 1/2 and so that $\langle m \rangle_{m \given m_\infty} = \langle m^{-1} \rangle_{m \given m_\infty}$.

\section{Comparison to simpler models}
\begin{figure}
\includegraphics[width=\textwidth]{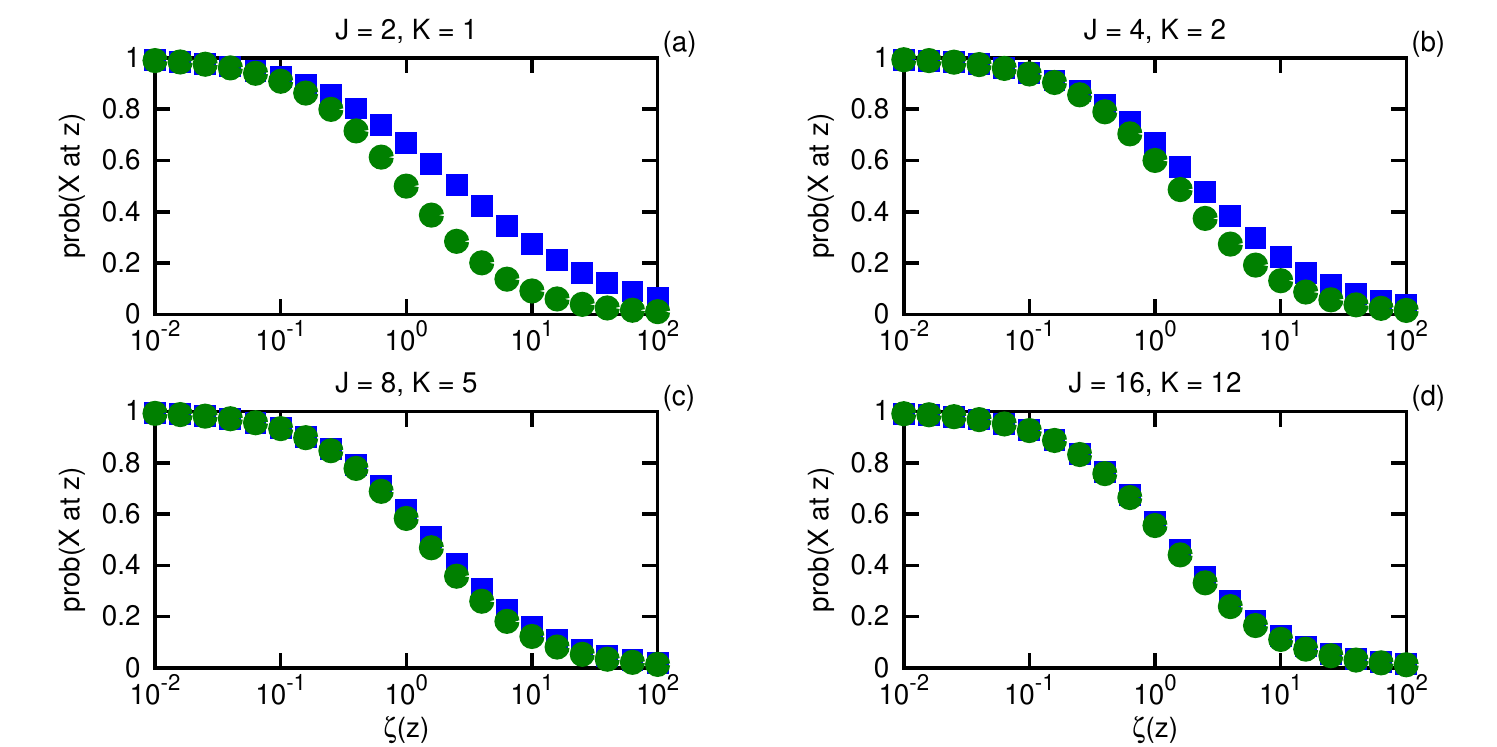}
\caption[]{Comparison of prob($X$ at $z$) as a function of $\zeta(z)$ estimated from the expectation value $\langle \p{X \mathrm{\,at\,} z}{z, \zeta, m} \rangle_{m \given J, K}$ shown as $\square$ to its approximation $\p{X \mathrm{\,at\,} z}{z, \zeta, \langle m \rangle_{m \given J, K}}$ shown as $\bigcirc$ for $J$ and $K$ as indicated above each panel.}
\label{fig:A}
\end{figure}

If the parameters $\bar{f}$, $\bar{g}$, $\tilde{f}$, and $\tilde{g}$ for the Gaussian distributions are known in advance, then only $m$ need be estimated from the data.  Using the notation $\q{\alpha}{\beta} \equiv - \log \p{\alpha}{\beta}$, the ``parameter info'' (information content of the evidence density) is \beq
\q{m}{J, K} = \q{J, K}{m} + \q{m}{} + C = N \log (1+m) - (K-1) \log m + C \;,
\eeq where $C$ is the logarithm of the normalizing constant, equal to $\log \beta(J,K)$ when $m_\infty = \infty$.  Its mode may be found from the equation for a vanishing gradient \beq
0 = \dsub{m} \q{m}{J, K} = m^{-1} (1+m)^{-1} [ 1 - K + m (J + 1) ] \; ,
\eeq whose solution $m_0 = (K-1) / (J+1)$ may be negative when $K = 0$, in which case the mode is at the lower limit of consideration; if one inverts the identity of $X$ and $Y$, one finds that the evidence for the inverted $m$ has its mode at the position given by the analytic formula.  In the limit of $m \in [0, \infty]$, the expectation that some new particle is of type $X$ is $\langle \p{X}{m} \rangle_{m \given J, K} = (1 + K/J)^{-1}$, the expectation value for $m$ is $\langle m \rangle_{m \given J, K} = K / (J - 1)$, and that for $m^{-1}$ is $\langle m^{-1} \rangle_{m \given J, K} = J / (K - 1)$.  When the new particle's location $z$ also is known, one may evaluate $\langle \p{X \mathrm{\,at\,} z}{z, \zeta, m} \rangle_{m \given J, K}$ as the prediction for its type being $X$; when both $J$ and $K$ are large, that value is approximately given by simply using the expectation for $m$ in the model for $\p{X \mathrm{\,at\,} z}{z, \zeta, m}$, \textit{i.e.} $\langle \p{X \mathrm{\,at\,} z}{z, \zeta, m} \rangle_{m \given J, K} \approx \p{X \mathrm{\,at\,} z}{z, \zeta, \langle m \rangle_{m \given J, K}}$ for $J, K \gg 1$.  In Figure~\ref{fig:A} we compare the value of $\langle \p{X \mathrm{\,at\,} z}{z, \zeta, m} \rangle_{m \given J, K}$ to $\p{X \mathrm{\,at\,} z}{z, \zeta, \langle m \rangle_{m \given J, K}}$ as a function of $\zeta(z)$ for various values of $J$ and $K$.

Let us next consider the case where only the deviations $\tilde{f}$ and $\tilde{g}$ are known in advance, so that now $\bar{f}$ and $\bar{g}$ must also be estimated from the data.  For convenience, let us further suppose that $\tilde{f} = \tilde{g} = 1$, setting the unit for the locations $z$.  The values of the parameters used to generate the data will be denoted $m_\Omega$, $\bar{f}_\Omega$, and $\bar{g}_\Omega$, where the subscript $\Omega$ indicates conditioning on the sum of all knowledge, \textit{i.e.} they are the ``true'' values unknown to mere mortals.  According to the model, the actual chance of finding an $X$ at some $z$ is given by $\p{X \mathrm{\,at\,} z}{z, \Omega} = [1 + m_\Omega \zeta_\Omega(z)]^{-1}$, thus the estimation of that quantity, as well as its reliability, from the data at hand is the desired goal of the statistical analysis.  The manifold position $\mbf{r}_\Omega$ of course is not allowed to be part of that process, as its knowledge would preclude the need to collect any data for its estimation.

Retaining only $m$, $\bar{f}$, and $\bar{g}$ in $\mbf{r}$, the evidence density is \beq
\p{\mbf{r}}{\mbf{X}, \mbf{Y}} \propto \p{J, K}{m} \p{\mbf{X}}{\bar{f}} \p{\mbf{Y}}{\bar{g}} \p{m, \bar{f}, \bar{g}}{} \; ,
\eeq where the prior is normalized to unit volume, thus the parameter info is \beq
\q{\mbf{r}}{\mbf{X}, \mbf{Y}} = \q{J, K}{m} + \q{m}{} + \q{\mbf{X}}{\bar{f}} + \q{\mbf{Y}}{\bar{g}} + C \; ,
\eeq and the value of $C$ is chosen according to the task at hand.  For taking expectation values, one convenient choice is that which normalizes the peak of the evidence to unity $C_0 = -q_0$, whereas for hypothesis testing (model selection) $C$ must equal the logarithm of the normalizing constant for the prior, and the remaining terms must retain any constants found in the likelihood, unless they happen to cancel out of the relative evidence ratio.  Here, $C_\mbf{r} = \log (2 \log m_\infty) + 2 \log \Delta_z$ when $m \in [m_\infty^{-1}, m_\infty]$ and $\bar{f}, \bar{g} \in [-\Delta_z/2, \Delta_z/2]$.  The expression \beq
\q{\mbf{X}}{\bar{f}} + \q{\mbf{Y}}{\bar{g}} = \dfrac{N}{2} \log (2 \pi) + \dfrac{1}{2} \sum_{z \in \mbf{X}} ( \bar{f} - z)^2 + \dfrac{1}{2} \sum_{z \in \mbf{Y}} ( \bar{g} - z)^2 
\eeq gives the additional likelihood info coming from the location measurements, and its mode is easily found to be at $\bar{f}_0 = \langle X_j \rangle_j$ and $\bar{g}_0 = \langle Y_k \rangle_k$.

Because the preferred locations $\bar{f}$ and $\bar{g}$ are themselves estimated, one must ask whether the data would be more efficiently represented by a single Gaussian, which we will call $h(z)$ with mean $\bar{h}$.  The relevant factors in the relative evidence ratio are those which do not depend on $m$, and the problem reduces to the well-known example of whether the difference in means between two populations is statistically significant.  The answer is given by the ratio of the expectation values for the likelihood of each model, \beq
\rho^{f g}_h \equiv \dfrac{\langle \p{\mbf{X}, \mbf{Y}}{\bar{f}, \bar{g}} \rangle_{\bar{f}, \bar{g}}}{\langle \p{\mbf{X}, \mbf{Y}}{\bar{h}} \rangle_{\bar{h}}} = \dfrac{\int_{-\Delta_z/2}^{\Delta_z/2} \int_{-\Delta_z/2}^{\Delta_z/2} \p{\mbf{X}, \mbf{Y}}{\bar{f}, \bar{g}} \p{\bar{f}, \bar{g}}{} \, d\bar{f} d\bar{g}}{\int_{-\Delta_z/2}^{\Delta_z/2} \p{\mbf{X}, \mbf{Y}}{\bar{h}} \p{\bar{h}}{} \, d\bar{h}} \approx \dfrac{\p{\mbf{X}}{\bar{f}_0} \p{\mbf{Y}}{\bar{g}_0}}{\p{\mbf{X}, \mbf{Y}}{\bar{h}_0}} \left( \dfrac{2 \pi N}{\Delta_z^2 J K} \right)^{1/2} \; ,
\eeq where the first factor is the ratio of peak likelihoods (when the prior is uniform), the second (Occam) factor is the ratio of the filling fractions for each model, and the approximation results from taking infinite bounds in the integrals.  The filling fraction for a model is a number between 0 and 1 which indicates how much the evidence fills the parameter manifold with respect to the prior measure. If the relative evidence ratio is well below unity $\rho^{f g}_h \ll 1$, then probability theory is telling one to neglect the $z$ dependence and simply use $m$ as the basis for any prediction regarding the type of some unidentified particle.  If either $J$ or $K$ equals 0, attention to factors reveals that $\rho^{f g}_h = 1$ in that case, as the evidence density is uniform along the irrelevant parameter.

\begin{figure}
\includegraphics[width=\textwidth]{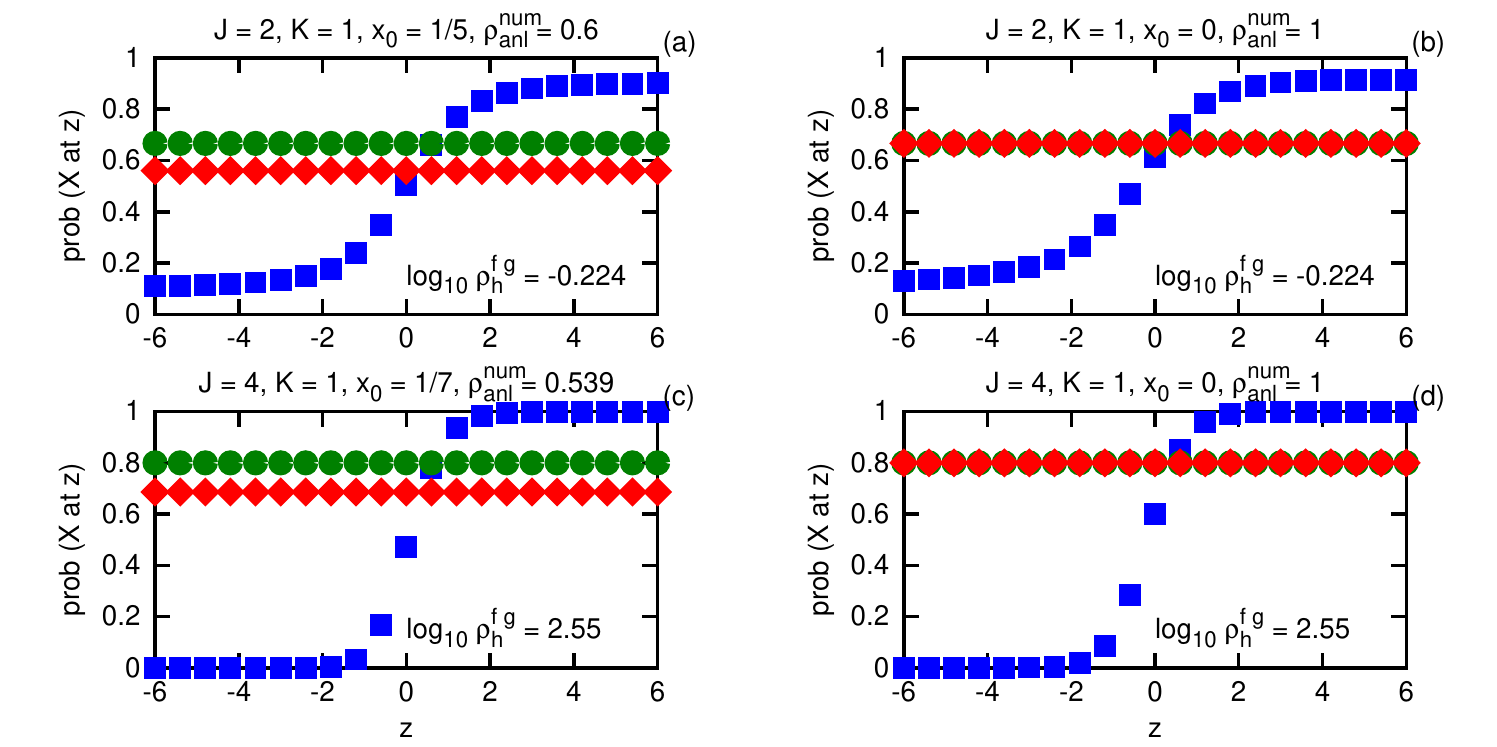}
\caption[]{Comparison of the numerical estimate for $\langle \p{X}{m} \rangle_{m \given J, K}$ displayed as $\lozenge$ to that for $\langle \p{X \mathrm{\,at\,} z}{z, \mbf{r}} \rangle_{\mbf{r} \given \mbf{X}, \mbf{Y}}$ displayed as $\square$ using limits of integration given by $\Delta_z = 12$ and $x_0$ as indicated above each panel, with the analytic value for $\langle \p{X}{m} \rangle_{m \given J, K}$ displayed as $\bigcirc$.}
\label{fig:B}
\end{figure}

Using finite limits for numerical integration can have an impact on the result.  Recalling that conditioning on $N$ has been implied throughout, let us compare the estimate of $\langle \p{X}{m} \rangle_{m \given J, K}$ to $\langle \p{X \mathrm{\,at\,} z}{z, \mbf{r}} \rangle_{\mbf{r} \given \mbf{X}, \mbf{Y}}$ for the two cases of $m_\infty = N+1$ and $m_\infty = \infty$.  The former represents a prior state of knowledge in which one is certain to observe particles of both types eventually, even though particles of only one type have been observed so far.  The numerical integration is more easily accomplished upon a change of variables $x = (1+m)^{-1}$ such that \beq
\int_{1/m_\infty}^{m_\infty} m^{K-1} (1+m)^{-J-K} \, dm = \int_{x_0}^{1-x_0} x^{J-1} (1-x)^{K-1} \, dx \leq \beta(J,K) \;,
\eeq with equality in the limit $x_0 \rightarrow 0$.  One way to assess the reliability of one's estimate is to inspect the ratio of the numerically integrated evidence volume to that derived analytically for the infinite manifold, identified as $\rho^\mathrm{num}_\mathrm{anl}$.  Figure~\ref{fig:B} shows the comparison of the estimated operators for values of $N = 3$ and $N = 5$ using $\bar{f}_\Omega = 1$ and $\bar{g}_\Omega = -1$, as well as the analytic expression for $\langle \p{X}{m} \rangle_{m \given J, K}$ when $m_\infty = \infty$.  With only a few measurements, the  numerical estimate for $\langle \p{X \mathrm{\,at\,} z}{z, \mbf{r}} \rangle_{\mbf{r} \given \mbf{X}, \mbf{Y}}$ approaches the analytic value for $\langle \p{X}{m} \rangle_{m \given J, K}$ only when $m_\infty$ is very large.  After more measurements have accumulated, the limits on $m$ have less impact; however, if the mode in $m$ is close to the numerical limit, the estimate may still be inaccurate.

\begin{figure}
\includegraphics[width=\textwidth]{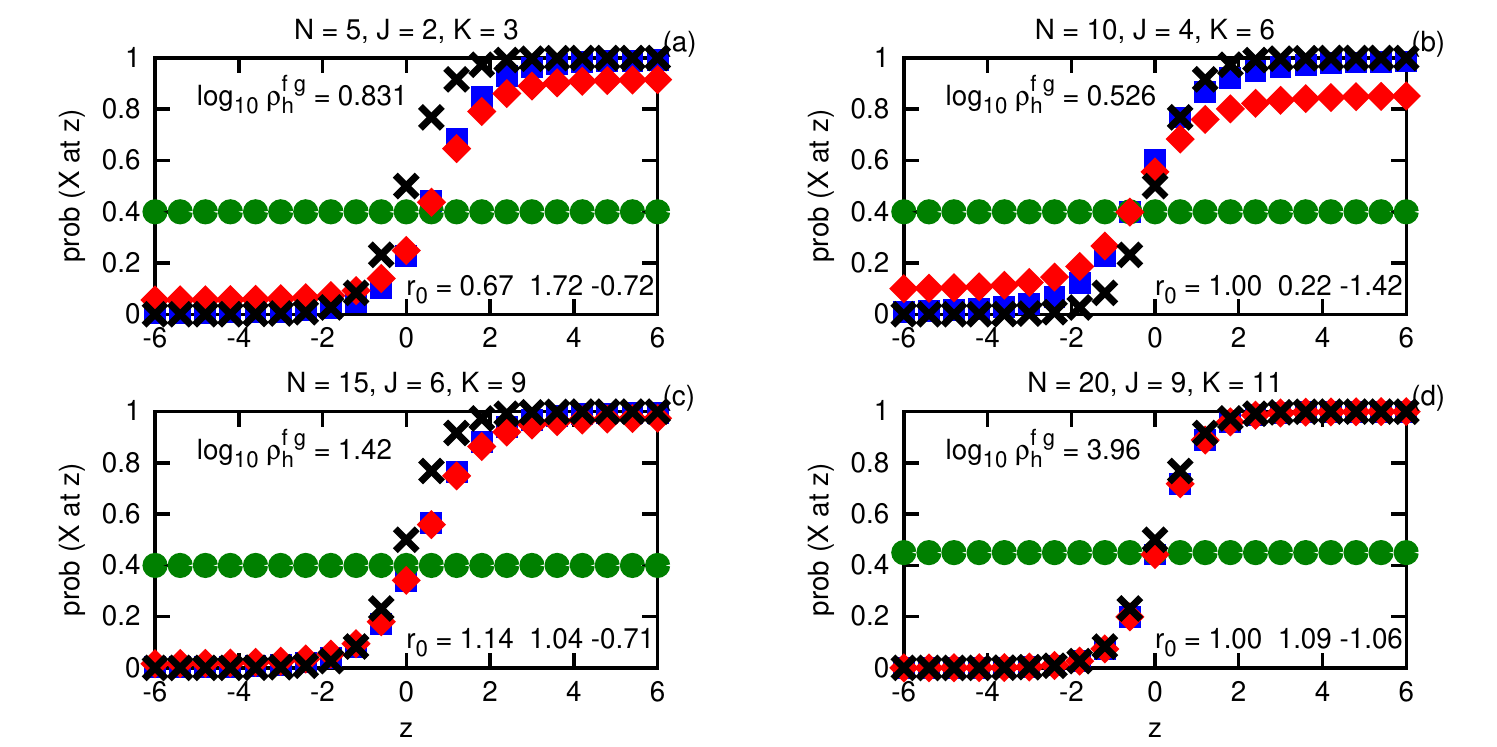}
\caption[]{Comparison of the analytic value for $\langle \p{X}{m} \rangle_{m \given J, K}$ displayed as $\bigcirc$ to the numerical estimate for $\langle \p{X \mathrm{\,at\,} z}{z, \mbf{r}} \rangle_{\mbf{r} \given \mbf{X}, \mbf{Y}}$ displayed as $\square$ for values of $N$ as indicated above each panel, with their weighted mean $\p{X \mathrm{\,at\,} z}{z, \rho}$ displayed as $\lozenge$ and the underlying distribution $\p{X \mathrm{\,at\,} z}{z, \Omega}$ displayed as $\times$.  Also shown are the values for the parameter mode $\mbf{r}_0$.}
\label{fig:C}
\end{figure}

To collect the estimates from the two possible models into a single prediction, one simply averages them with weights given by their relative evidence, \beq
\p{X \mathrm{\,at\,} z}{z, \rho} \equiv \left[ \rho^{f g}_h \langle \p{X \mathrm{\,at\,} z}{z, \mbf{r}} \rangle_{\mbf{r} \given \mbf{X}, \mbf{Y}} + \langle \p{X}{m} \rangle_{m \given J, K} \right] \left( \rho^{f g}_h + 1 \right)^{-1} \; .
\eeq  As evidence accumulates in favor of the $f g$ model, their average quickly approaches the estimate from just that model, as seen in Figure~\ref{fig:C}.  Using values of $m_\Omega = 1$, $\bar{f}_\Omega = -\bar{g}_\Omega = 1$ and integration limits of $\Delta_z = 12$ and $x_0 = 0$, it takes only a few tens of measurements before the evidence for the null hypothesis (the $h$ model) is negligible.  As the number of measurements in the training data grows, the estimate $\p{X \mathrm{\,at\,} z}{z, \rho}$ converges to the underlying $\Omega$ distribution for the chance of finding an $X$ at $z$.

So far we have said nothing about the rate of convergence of the estimate $\langle \p{X \mathrm{\,at\,} z}{z, \mbf{r}} \rangle_{\mbf{r} \given \mbf{X}, \mbf{Y}}$ as a function of the number of measurements $N$.  Partly that is because we have summarized our inference about $\p{X \mathrm{\,at\,} z}{z, \mbf{X}, \mbf{Y}}$ into a single number, the expectation value of the observable $\p{X \mathrm{\,at\,} z}{z, \mbf{r}}$, and to answer the question of convergence requires keeping track of two numbers for the inference, representing for example its central location and its width.  That procedure will be addressed later in this article.  While we have blithely displayed $\p{X \mathrm{\,at\,} z}{z, \Omega}$ in the preceding figure, one should never forget that its knowledge is beyond the ken of mortals.  Practically speaking, one simply must collect a sufficient amount of data such that collecting more data no longer significantly influences the estimate, implicitly assuming that the underlying physical process is stationary in time.

\begin{figure}
\includegraphics[width=\textwidth]{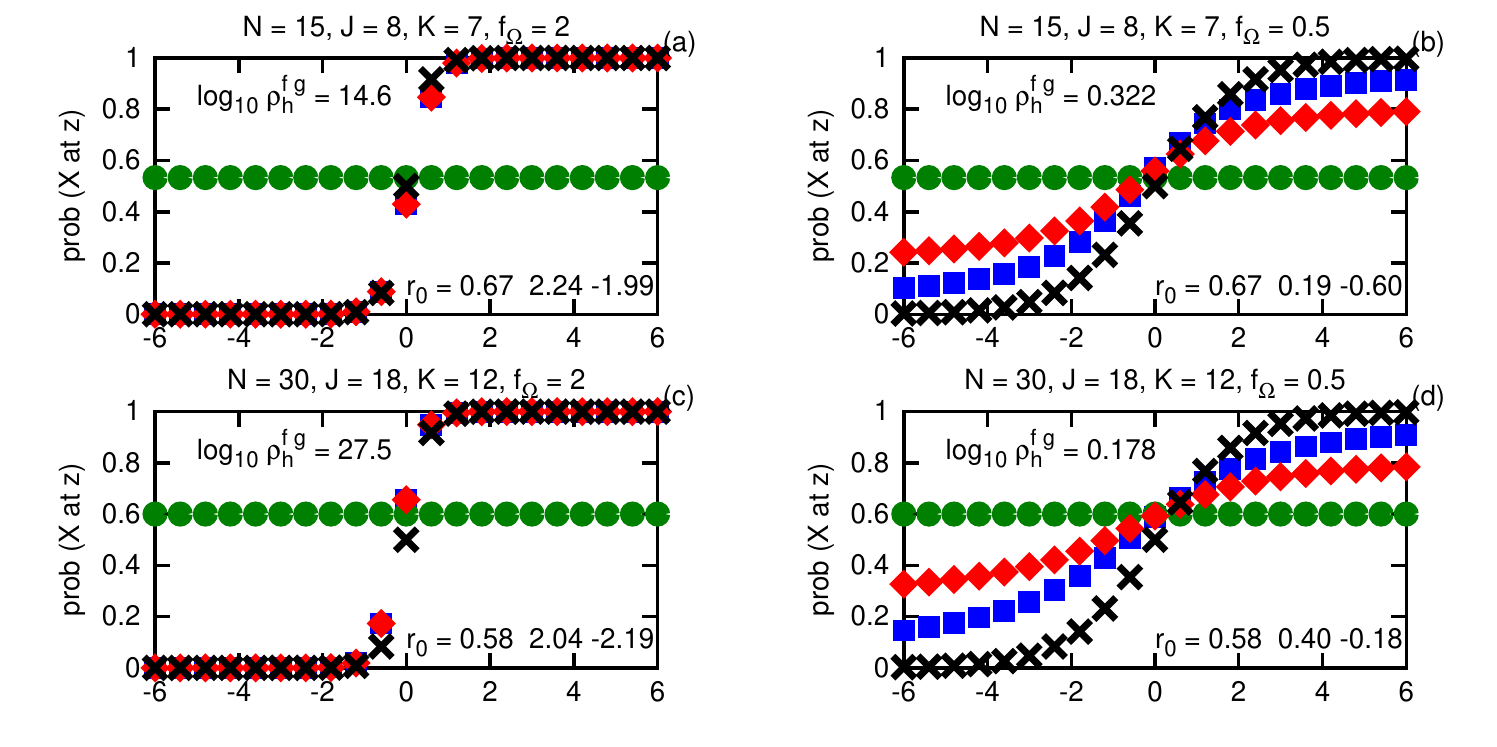}
\caption[]{Comparison of the analytic value for $\langle \p{X}{m} \rangle_{m \given J, K}$ displayed as $\bigcirc$ to the numerical estimate for $\langle \p{X \mathrm{\,at\,} z}{z, \mbf{r}} \rangle_{\mbf{r} \given \mbf{X}, \mbf{Y}}$ displayed as $\square$ for values of $N$ and $\bar{f}_\Omega = -\bar{g}_\Omega$ as indicated above each panel, with their weighted mean $\p{X \mathrm{\,at\,} z}{z, \rho}$ displayed as $\lozenge$ and the underlying distribution $\p{X \mathrm{\,at\,} z}{z, \Omega}$ displayed as $\times$.  Also shown are the values for the parameter mode $\mbf{r}_0$.}
\label{fig:D}
\end{figure}

The ratio of the the deviation in the data $\tilde{f}$ to the separation of the preferred locations $\bar{f} - \bar{g}$ affects how much data is necessary for convergence of the estimate.  When that ratio is small, not many measurements are needed before the null hypothesis is discounted, and further measurements serve only to improve the convergence.  However, when that ratio is large, the null hypothesis can be discounted only after a sufficient number of measurements have been taken so that the parameter evidence is well resolved.  In Figure~\ref{fig:D} we show the same estimates as in Figure~\ref{fig:C} but for $N$ equal to 15 and 30 and $\bar{f}_\Omega = -\bar{g}_\Omega$ of 2 and 1/2.  Even when the mode $\mbf{r}_0$ gives a poor reckoning of the underlying process, the expectation value $\langle \p{X \mathrm{\,at\,} z}{z, \mbf{r}} \rangle_{\mbf{r} \given \mbf{X}, \mbf{Y}}$ is fairly accurate when $\rho^{f g}_h \gg 1$.  To make from $\rho^{f g}_h$ a number comparable to the $P$ (or $Q$) value of frequentist methods, one would state that the null hypothesis ($h$ model) is discounted at the level of $(\rho^{f g}_h + 1)^{-1}$.

\section{Integration over unknown deviation}
Let us now consider the more realistic situation where the deviations of the location distributions are not known in advance.  For convenience, let us assign them all the domain of $\sigma \in [\sigma_0, \sigma_\infty]$ with prior $\p{\sigma}{} \propto \sigma^{-1}$.  The normalization constant is $\Delta_{\log \sigma} \equiv \log \sigma_\infty - \log \sigma_0$, which equals infinity if either $\sigma_\infty = \infty$ or $\sigma_0 = 0$.  The first task is to evaluate the relative evidence of the models, \beq
\rho^{f g}_h = \dfrac{\int_{-\Delta_z/2}^{\Delta_z/2} \int_{-\Delta_z/2}^{\Delta_z/2} \int_{\sigma_0}^{\sigma_\infty} \int_{\sigma_0}^{\sigma_\infty} \tilde{f}^{-J-1} \tilde{g}^{-K-1} \exp^{-1/2} ( \chi^2_f + \chi^2_g ) \, d\tilde{f} d\tilde{g} d\bar{f} d\bar{g}}{\Delta_z \Delta_{\log \sigma} \int_{-\Delta_z/2}^{\Delta_z/2} \int_{\sigma_0}^{\sigma_\infty} \tilde{h}^{-N-1} \exp^{-1/2} ( \chi^2_h ) \, d\tilde{h} d\bar{h}} \; ,
\eeq where $\chi^2_f \equiv \sum_{z \in \mbf{X}} [ (\bar{f} - z) / \tilde{f} ]^2$ and similarly for $\chi^2_g$ and $\chi^2_h$, which can be approximated as before by taking the Gaussian integrals over infinite limits yet retaining the finite normalization to yield \beq
\rho^{f g}_h \approx \left( \dfrac{2 \pi N}{\Delta_z^2 J K} \right)^{1/2} \dfrac{\int_{\sigma_0}^{\sigma_\infty} \int_{\sigma_0}^{\sigma_\infty} \tilde{f}^{-J} \tilde{g}^{-K} \exp^{-1/2} ( \xi^2_f / \tilde{f}^2 + \xi^2_g / \tilde{g}^2 ) \, d\tilde{f} d\tilde{g}}{\Delta_{\log \sigma} \int_{\sigma_0}^{\sigma_\infty} \tilde{h}^{-N} \exp^{-1/2} ( \xi^2_h / \tilde{h}^2 ) \, d\tilde{h}} \; ,
\eeq where $\xi_f^2 \equiv J ( \langle X_j^2 \rangle_j - \langle X_j \rangle_j^2 )$ and similarly for $\xi_g^2$ and $\xi_h^2$.  The remaining integrals can be evaluated analytically to give the result \beq
\rho^{f g}_h \approx \left( \dfrac{\pi N}{J K} \right)^{1/2} \dfrac{ \xi_h^{N-1} \Delta_\Gamma (\xi_f) \Delta_\Gamma (\xi_g)}{2 \Delta_z \Delta_{\log \sigma} \, \xi_f^{J-1} \xi_g^{K-1} \Delta_\Gamma (\xi_h)} \; ,
\eeq where $\Delta_\Gamma$ is defined in terms of the upper incomplete gamma function $\Gamma (a,z) \equiv \int_z^\infty t^{a-1} e^{-t} dt$, such that \beq
\Delta_\Gamma (\xi_f) \equiv \Gamma \left( \dfrac{J-1}{2} , \dfrac{\xi_f^2}{2 \sigma_\infty^2} \right) - \Gamma \left( \dfrac{J-1}{2} , \dfrac{\xi_f^2}{2 \sigma_0^2} \right) \; ,
\eeq and similarly for $\Delta_\Gamma (\xi_g)$ and $\Delta_\Gamma (\xi_h)$.  Note that this formulation makes no use of the peak evidence ratio but instead is expressed entirely in terms of the data and the limits of the prior.

To recover the peak evidence ratio (and thus the Occam factor), one needs to evaluate the model evidence densities at their mode positions.  The parameter info for the $\mbf{X}$ data is now \beq
\q{\bar{f}, \tilde{f}}{\mbf{X}} = ( J + 1 ) \log \tilde{f} + (2 \tilde{f}^2)^{-1} \sum_{z \in \mbf{X}} ( \bar{f} - z )^2 + C \; ,
\eeq whose gradient is given by \beq
\del \q{\bar{f}, \tilde{f}}{\mbf{X}} = \tilde{f}^{-3} \left[ \begin{array}{c} 
\tilde{f} \sum_{z \in \mbf{X}} ( \bar{f} - z ) \\ 
( J + 1 ) \tilde{f}^2 - \sum_{z \in \mbf{X}} ( \bar{f} - z )^2 
\end{array} \right] \; ,
\eeq which vanishes at the mode $\del \q{\bar{f}_0, \tilde{f}_0}{\mbf{X}} = 0$.  As before, the mode in $\bar{f}$ is at the mean of the locations $\bar{f}_0 = \langle X_j \rangle_j$, and the mode in $\tilde{f}$ can be written as $\tilde{f}_0 = (J+1)^{-1/2} \xi_f$.  The position of the mode for the remaining Gaussian parameters is found similarly.

Let us briefly discuss the limits of integration hence the normalization of the prior.  If the prior is not to be based upon the current crop of measurements, where does the information for the limits come from?  The practical answer is that the limits are determined by the nature of the measurement apparatus.  Any set of measurements collected within a finite span of time necessarily are limited by the range and resolution of the device used for their collection, for example measurements of the voltage of a circuit collected by a common voltmeter.  As long as no measurement ``pegs the needle'' one can safely use limits based on the range of the device; those that do can be addressed through an appropriately modified contribution to the likelihood beyond the scope of this article.  Similarly, the resolution of the device (or the width of the particle) sets a lower limit on what can be said about any measured deviations in the population locations.  For the evaluation of the evidence ratio $\rho^{f g}_h$ as well as $\rho^\mathrm{num}_\mathrm{anl}$, we will set the limits for the deviations as $\sigma_\infty = \Delta_z$ and $\sigma_0 = 10^{-4} \Delta_z$, with $\Delta_z = 12$ and $x_0 = 0$ as above.

The evaluation of the expectation value of the observable $\langle \p{X \mathrm{\,at\,} z}{z, \mbf{r}} \rangle_{\mbf{r} \given \mbf{X}, \mbf{Y}}$ proceeds as before, only now the integration is over a 5 dimensional parameter manifold.  As \textit{Numerical Recipes}~\cite{Press-1992} states, ``integrals of functions of several variables, over regions with dimension greater than one, are \textit{not easy}.''  For problems of Bayesian inference, the majority of the contribution to the integral comes from a region localized around the peak of the evidence density when sufficient data exists that the limits of the prior are irrelevant.  Luckily, for this problem the evidence mode is unique and analytic, so that one may select limits for the numerical integration much tighter than those given by the prior while still encompassing 99.9\% of the normalized evidence density.  The evaluation is performed using an adaptive grid algorithm~\cite{berntsen-1991437,genz-1980295} over a small region of the manifold centered on the position of the optimal parameter values.

\begin{figure}
\includegraphics[width=\textwidth]{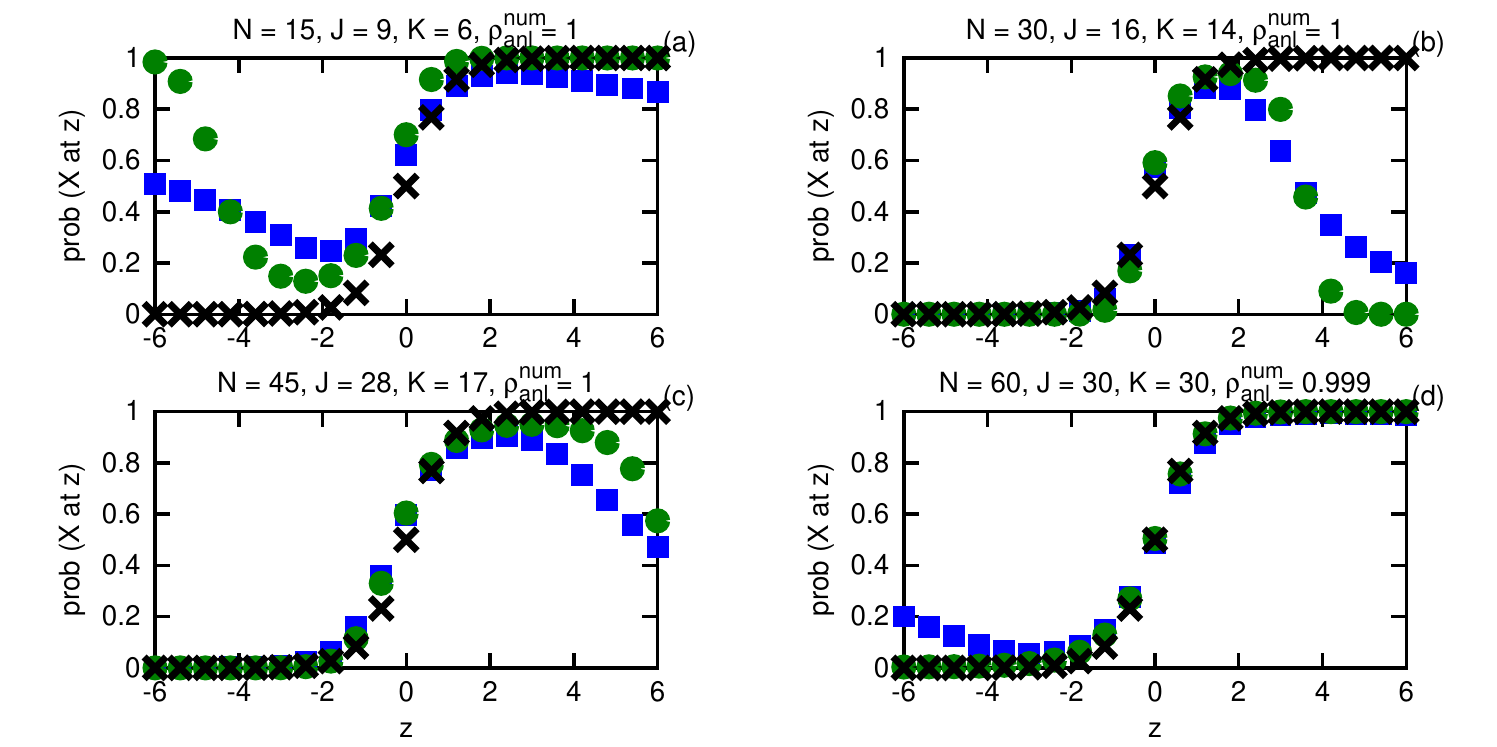}
\caption[]{Comparison of the expectation value $\langle \p{X \mathrm{\,at\,} z}{z, \mbf{r}} \rangle_{\mbf{r} \given \mbf{X}, \mbf{Y}}$ displayed as $\square$ to the estimate from the parameter mode $\p{X \mathrm{\,at\,} z}{z, \mbf{r}_0 \given \mbf{X}, \mbf{Y}}$ displayed as $\bigcirc$ for $N = 50$ as well as the underlying distribution $\p{X \mathrm{\,at\,} z}{z, \Omega}$ displayed as $\times$ with parameter values as given in Table~\ref{tab:A}.}
\label{fig:E}
\end{figure}

\begin{table}
\caption{\label{tab:A} Parameters corresponding to Figure~\ref{fig:E}}
\centering
\fbox{%
\begin{tabular}{c|rrrrrrr} 
\multirow{2}{*}{panel} & $\log_{10} \rho^{f g}_h$ & $N$ & $m_\Omega$ & $\bar{f}_\Omega$ & $\bar{g}_\Omega$ &  $\tilde{f}_\Omega$ & $\tilde{g}_\Omega$ \\
 & $J$ &  $K$ & $m_0$ & $f_0$ & $g_0$ & $\tilde{f}_0$ & $\tilde{g}_0$ \\ \hline
\multirow{2}{*}{a} & -0.397 &    15 &     1 &     1 &    -1 &     1 &     1 \\       
                   &     9 &     6 &   0.5 & 0.634 & -0.84 & 0.994 & 0.715  \\ \hline
\multirow{2}{*}{b} &   3.1 &    30 &     1 &     1 &    -1 &     1 &     1  \\       
                   &    16 &    14 & 0.765 &  0.87 & -1.05 & 0.654 &  1.19  \\ \hline
\multirow{2}{*}{c} &  3.45 &    45 &     1 &     1 &    -1 &     1 &     1  \\       
                   &    28 &    17 & 0.552 &  1.15 & -0.85 & 0.953 &  1.38  \\ \hline
\multirow{2}{*}{d} &  4.77 &    60 &     1 &     1 &    -1 &     1 &     1  \\       
                   &    30 &    30 & 0.935 &   0.9 & -0.879 &  1.08 & 0.936 \\       
\end{tabular}}
\end{table}

In Figure~\ref{fig:E} we compare the expectation of the observable $\langle \p{X \mathrm{\,at\,} z}{z, \mbf{r}} \rangle_{\mbf{r} \given \mbf{X}, \mbf{Y}}$ to that given by evaluating the observable using the parameter mode $\p{X \mathrm{\,at\,} z}{z, \mbf{r}_0 \given \mbf{X}, \mbf{Y}}$, with the results given in Table~\ref{tab:A}.  As the number of measurements increases, those estimates draw closer, according to the narrowing of the peak in the evidence density.  The estimate from the expectation value is ``more conservative'' than that from the mode, in that it is closer to the estimate $\langle \p{X}{m} \rangle_{m \given \mbf{X}, \mbf{Y}}$ (not shown).  While the estimate from the mode more closely resembles the underlying distribution $\p{X \mathrm{\,at\,} z}{z, \Omega}$ when there is sufficient data, in a real world situation we are not privy to that knowledge (which would obviate the need for a statistical analysis).  The expectation value $\langle \p{X \mathrm{\,at\,} z}{z, \mbf{r}} \rangle_{\mbf{r} \given \mbf{X}, \mbf{Y}}$ summarizes what the available data have to say about the observable into a single number.  Using $\p{X \mathrm{\,at\,} z}{z, \mbf{r}_0 \given \mbf{X}, \mbf{Y}}$ as a proxy for the mode of the observable (which is not quite the same thing as the observable of the mode), one could ascribe to $\p{X \mathrm{\,at\,} z}{z, \mbf{r}}$ a beta distribution according to its mean and mode; the maximum likelihood estimate of the parameters of the beta distribution for the observable $\p{X \mathrm{\,at\,} z}{z, \mbf{r}}$ requires evaluation of the two observables $\langle \log \p{X \mathrm{\,at\,} z}{z, \mbf{r}} \rangle_{\mbf{r} \given \mbf{X}, \mbf{Y}}$ and $\langle \log ( 1 - \p{X \mathrm{\,at\,} z}{z, \mbf{r}} ) \rangle_{\mbf{r} \given \mbf{X}, \mbf{Y} }$.

\section{Comparison to non-parametric classification}
\begin{figure}
\includegraphics[width=\textwidth]{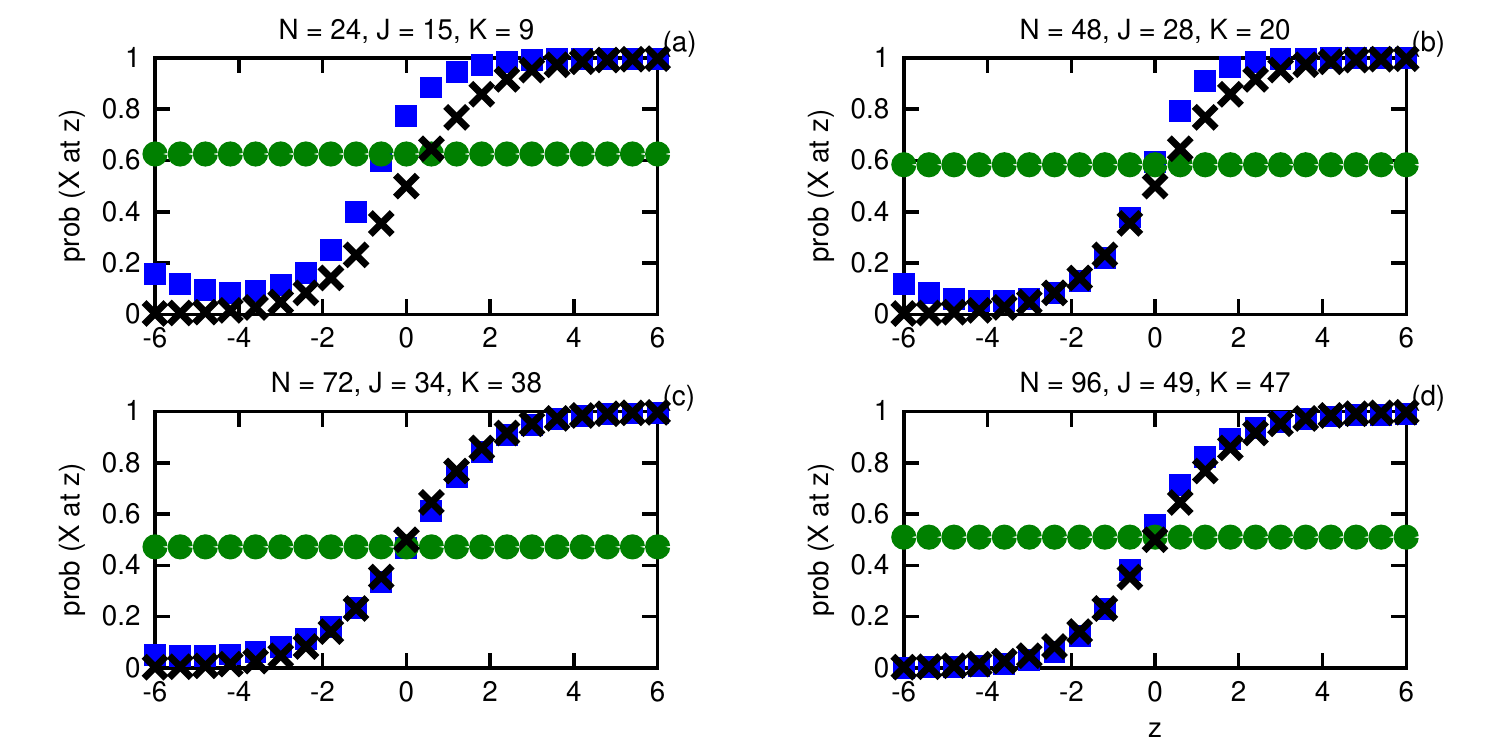}
\caption[]{Comparison of the expectation value $\langle \p{X \mathrm{\,at\,} z}{z, \mbf{r}} \rangle_{\mbf{r} \given \mbf{X}, \mbf{Y}}$ displayed as $\square$ to $\langle \p{X}{m} \rangle_{m \given \mbf{X}, \mbf{Y}}$ displayed as $\bigcirc$ for various $N$ as well as the underlying distribution $\p{X \mathrm{\,at\,} z}{z, \Omega}$ displayed as $\times$ with parameter values as given in Table~\ref{tab:D}.}
\label{fig:K}
\end{figure}

Let us now look at how the transformation group prediction compares to those derived from some non-parametric algorithms commonly employed for this type of problem.  For this section we will use the same data for each method generated using parameter values $m_\Omega = 1$, $\bar{f}_\Omega = -\bar{g}_\Omega = 2$, and $\tilde{f}_\Omega = \tilde{g}_\Omega = 2$ for various $N$.  We will consider both a nearest neighbor classification scheme which produces its estimate from a subset of the data ``close'' to the desired location as well as a classification scheme based upon a kernel density estimate of the identified particle distributions.  The transformation group estimates from these sets of data are shown in Figure~\ref{fig:K} with parameter modes in Table~\ref{tab:D}.

\begin{table}
\caption{\label{tab:D} Parameters corresponding to Figure~\ref{fig:K}}
\centering
\fbox{%
\begin{tabular}{c|rrrrrrr} 
\multirow{2}{*}{panel} & $\log_{10} \rho^{f g}_h$ & $N$ & $m_\Omega$ & $\bar{f}_\Omega$ & $\bar{g}_\Omega$ &  $\tilde{f}_\Omega$ & $\tilde{g}_\Omega$ \\
 & $J$ &  $K$ & $m_0$ & $f_0$ & $g_0$ & $\tilde{f}_0$ & $\tilde{g}_0$ \\ \hline
\multirow{2}{*}{a} &  3.09 &    24 &     1 &     2 &    -2 &     2 &     2 \\
 &    15 &     9 &   0.5 &  2.02 & -2.58 &  2.05 &  1.22 \\ \hline
\multirow{2}{*}{b} &  7.82 &    48 &     1 &     2 &    -2 &     2 &     2 \\
 &    28 &    20 & 0.655 &   2.3 & -1.99 &  1.98 &  1.27 \\ \hline
\multirow{2}{*}{c} &  6.77 &    72 &     1 &     2 &    -2 &     2 &     2 \\
 &    34 &    38 &  1.06 &  1.58 & -1.62 &  1.92 &  1.64 \\ \hline
\multirow{2}{*}{d} &  15.8 &    96 &     1 &     2 &    -2 &     2 &     2 \\
 &    49 &    47 &  0.92 &  2.08 & -2.52 &  1.76 &  2.11 \\ 
\end{tabular}}
\end{table}

Starting with nearest neighbor classification, its prediction for the type of some new particle, \beq
\p{X \mathrm{\,at\,} z}{z, \kappa, \mbf{X}, \mbf{Y}} \equiv \kappa_X / (\kappa_X + \kappa_Y) \; ,
\eeq is conditioned on the number of neighbors $\kappa = \kappa_X + \kappa_Y$ chosen to be influential, denoted here as $\kappa \equiv \rho_\kappa N$ for $0 < \rho_\kappa \leq 1$ such that $\kappa$ is an integer.  The selection of $\kappa$ is arbitrary, but Hall, \textit{et al.}~\cite{hall-5276H} describe a method for choosing its value based upon bootstrap estimates from the data.  Here, however, we are interested in the case where there is so little data that bootstrap estimates are unlikely to be reliable.  Consequently, we will consider the set of ratios $\rho_\kappa \in [1/8, 1/4, 1/2, 1]$.  Another distinction is that they assign a type of $X$ or $Y$ to the new particle at $z$ according to whether $\p{X \mathrm{\,at\,} z}{z, \kappa, \mbf{X}, \mbf{Y}}$ is greater or less than 1/2, such that the region boundaries form a decision surface in one dimension, rather than retaining the expression of the chance of finding an $X$ at $z$ as a probability.

\begin{figure}
\includegraphics[width=\textwidth]{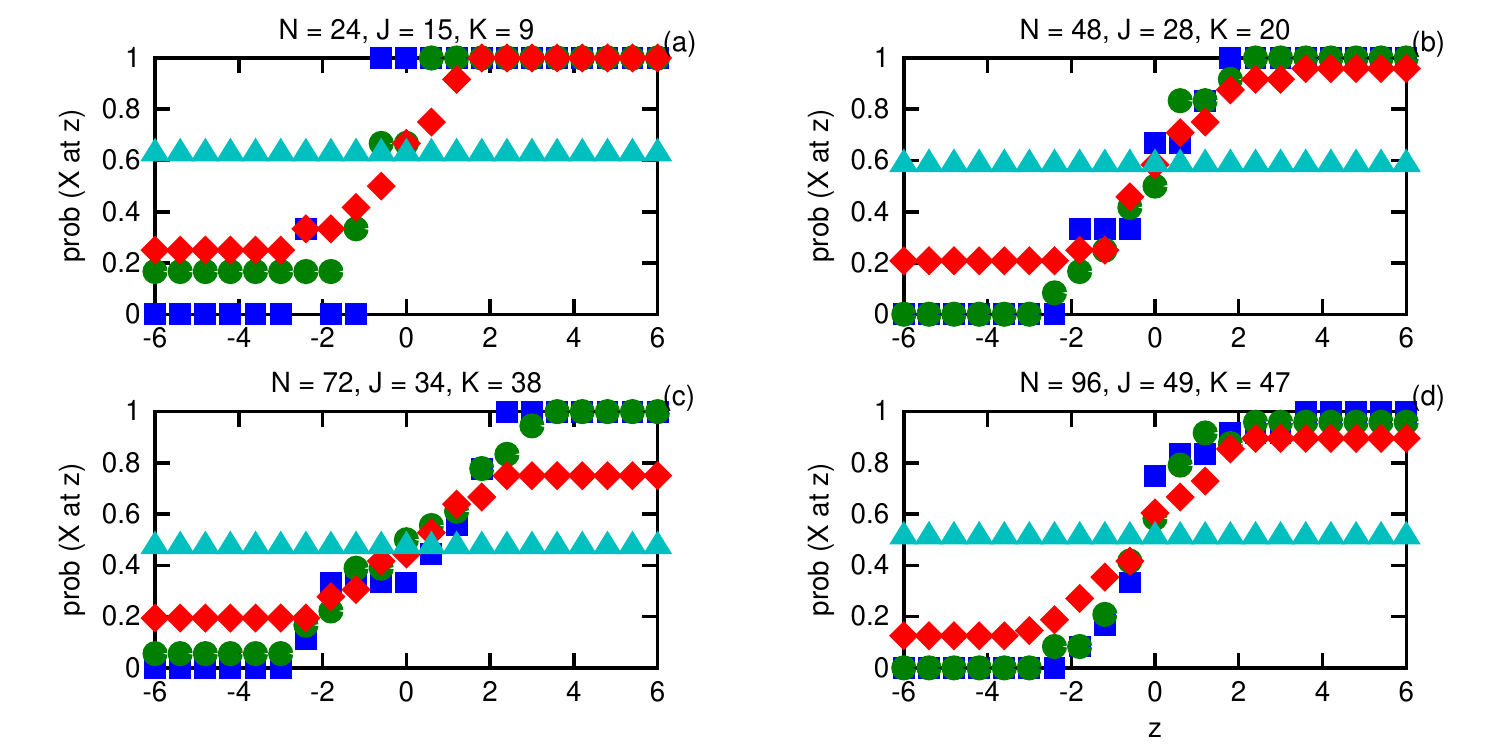}
\caption[]{Comparison of the nearest neighbor estimates $\p{X \mathrm{\,at\,} z}{z, \kappa, \mbf{X}, \mbf{Y}}$ for various $\kappa$ given by $\kappa / N \in [1/8, 1/4, 1/2, 1]$ shown as $\square$, $\bigcirc$, $\lozenge$, and $\triangle$ respectively.}
\label{fig:I}
\end{figure}

The results of the preceding method are shown in Figure~\ref{fig:I}.  One feature of the nearest neighbors method is that its prediction is quantized in units of $1 / \kappa$, which can lead to large jumps in the estimate when there is not much data.  These jumps yield an estimate which is not smooth as a function of $z$, even as the number of measurements approaches 100.  By basing its prediction on the rank of the distances from the data to the desired location, this method throws away information pertinent to the analysis.  Consequently, its prediction is a coarsely grained representation of the underlying distribution, even for moderately large sets of data.  If one were to implement a decision surface as in Hall, \textit{et al.}~\cite{hall-5276H}, one would have a prediction that oscillates wildly between 0 and 1 in the region where the particle types significantly overlap.

Alternately, we can consider a classification scheme based upon a kernel density estimate~\cite{terrell-1236,kim-2529,eberts-2013} of the particle distributions by type.  Its prediction $\p{X \mathrm{\,at\,} z}{z, \lambda, \mbf{X}, \mbf{Y}}$ is conditioned on the bandwidth parameter $\lambda$ for the kernel resolution.  The kernel basis chosen is that given by the Gaussian distributions $\delta_\lambda (z) \equiv \exp^{-1/2} (z^2 / \lambda^2)$, with peaks normalized to unity for convenience later.  The kernel density estimate for the distribution of type $X$ is equal to the sum of the kernel basis functions centered on the datum locations $f_{\lambda, \mbf{X}} (z) \equiv \sum_j \delta_\lambda (z-X_j)$, and similarly for $g_{\lambda, \mbf{Y}} (z)$.  The classification prediction is then determined from the ratio \beq
\p{X \mathrm{\,at\,} z}{z, \lambda, \mbf{X}, \mbf{Y}} \equiv f_{\lambda, \mbf{X}} (z) / [f_{\lambda, \mbf{X}} (z) + g_{\lambda, \mbf{Y}} (z) ] \; ,
\eeq conditioned on the value of the bandwidth parameter.  

\begin{figure}[]
\includegraphics[width=\textwidth]{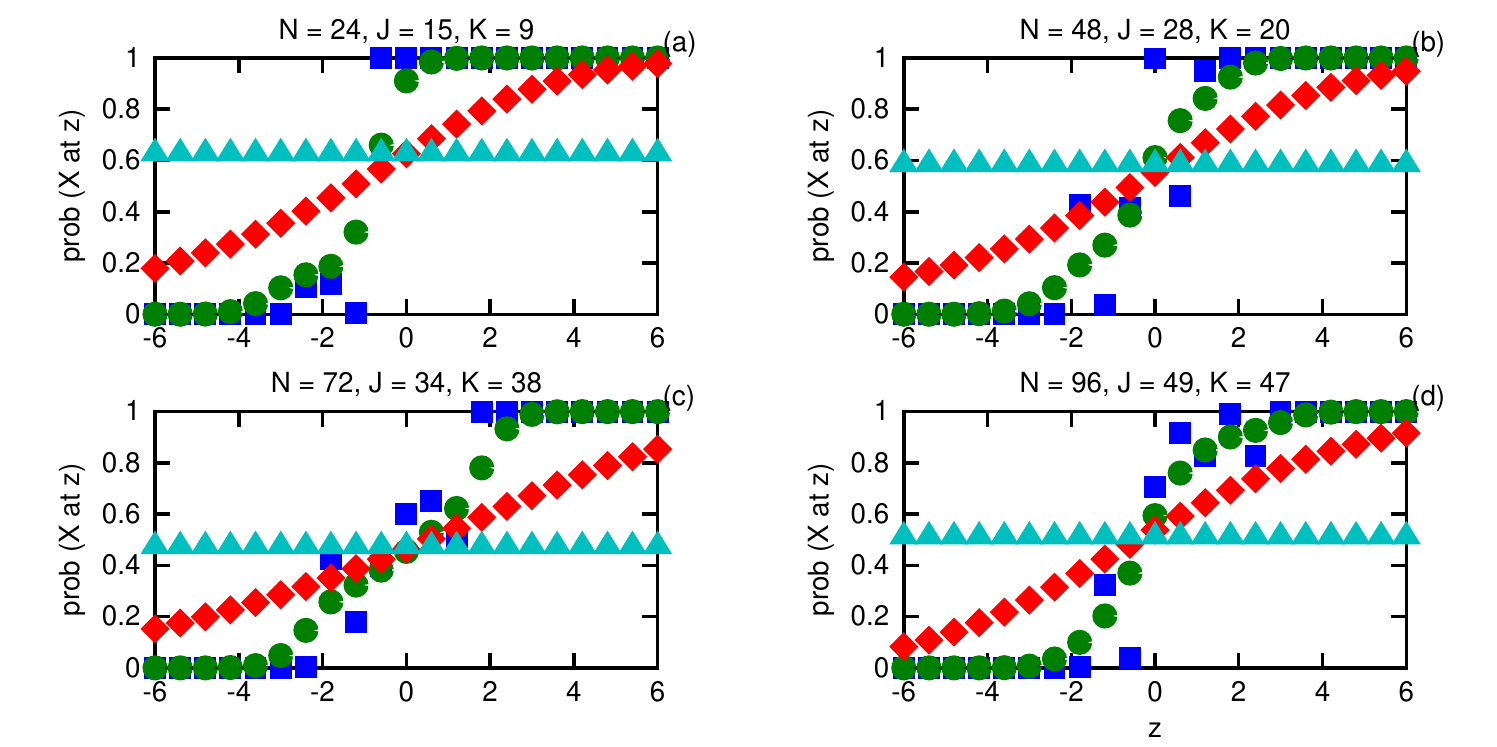}
\caption[]{Comparison of the kernel density estimate predictions $\p{X \mathrm{\,at\,} z}{z, \lambda, \mbf{X}, \mbf{Y}}$ for various $\lambda$ given by $\lambda \in [1/4, 1, 4, \infty]$ shown as $\square$, $\bigcirc$, $\lozenge$, and $\triangle$ respectively.}
\label{fig:J}
\end{figure}

The results for the preceding method, using values of $\lambda \in [1/4, 1, 4, \infty]$, are shown in Figure~\ref{fig:J}.  Again, large jumps are apparent in the estimate when the kernel width is much smaller than what we know the particle deviations to be, but the values are continuous rather than being quantized.  The effect of the kernel basis is that of a smoothing filter which spreads the information from each datum over a range of nearby locations according to $\lambda$.  In essence, with this method one is convoluting the data with some point spread function to produce an estimate of the observable at all locations $z$ based on the discrete list of measured locations for classified particles.

The term ``non-parametric'' is actually a bit of a misnomer, as the evaluation of either method requires specification of a parameter representing the bandwidth of the resolution filter.  While the bandwidth in $z$ for the kernel density prediction is constant for a given $\lambda$, for a given $\kappa$ that for the nearest neighbor prediction is not, based as it is on the rank of the distances in $z$ rather than their values.  In the limit of infinite bandwidth, such that the $z$ dependence disappears, both models give a prediction equal to that of the Bayesian estimate $\langle \p{X}{m} \rangle_{m \given \mbf{X}, \mbf{Y}} = J / (J+K)$, as indicated by $\triangle$ in the figures---the reason for the peak normalization of the kernel basis is so the kernel density estimates equal $J$ or $K$ for all $z$ in this limit.  The selection of the optimal value of the bandwidth parameter requires definition of some metric for its merit, introducing yet another source of subjectivity into the methodology.  In contrast, the only arbitrary elements of the transformation group method are the limits of the prior, everything else following from repeated applications of the rules of probability theory to the state of knowledge specified at the outset, and even those are not truly arbitrary when one considers the physical nature of the apparatus and the objects.

\section{Model mismatch}
\begin{figure}
\includegraphics[width=\textwidth]{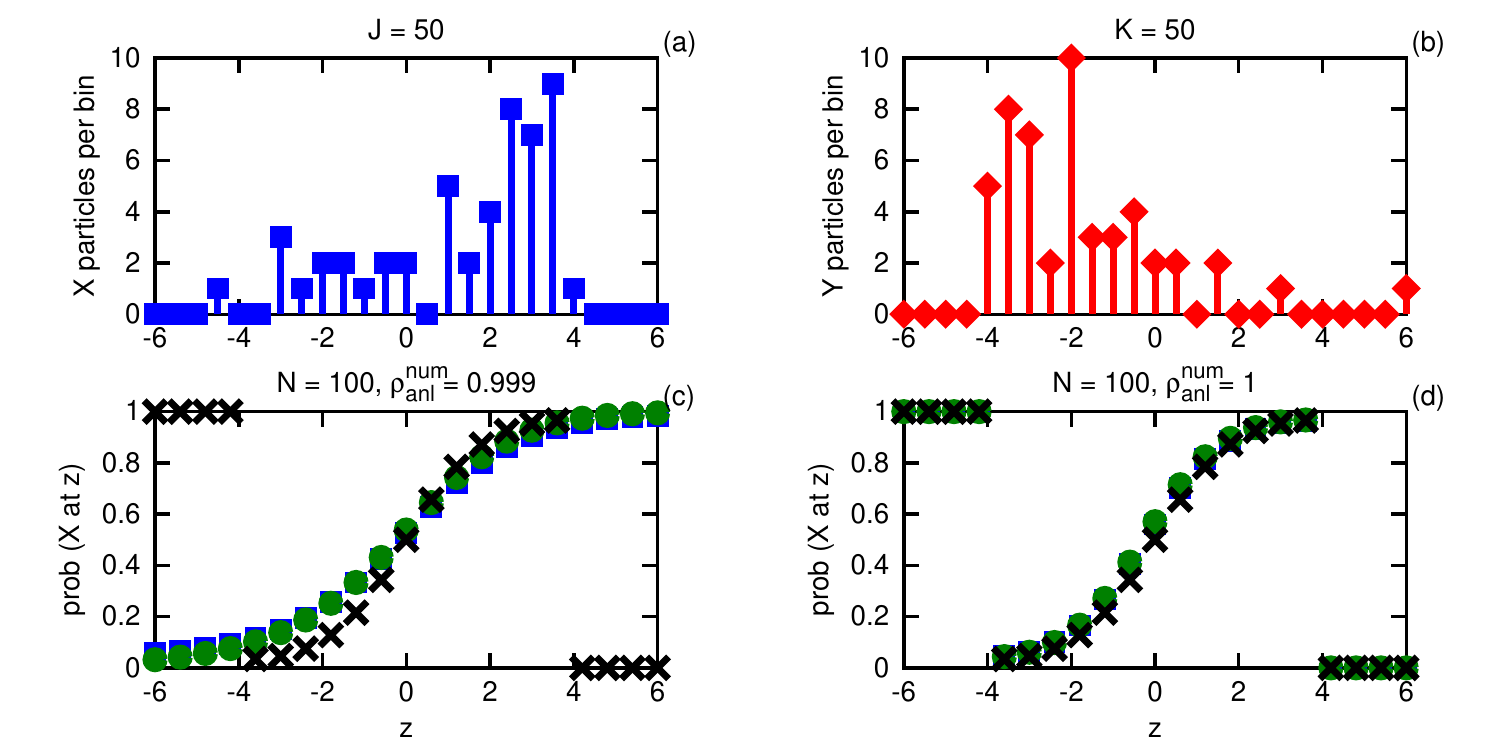}
\caption[]{Histograms of location for type $X$ in (a) and $Y$ in (b) with a bin width of 0.5 for a set of 100 particles which yield the predictions $\langle \p{X \mathrm{\,at\,} z}{z, \mbf{r}} \rangle_{\mbf{r} \given \mbf{X}, \mbf{Y}}$ displayed as $\square$ and $\p{X \mathrm{\,at\,} z}{z, \mbf{r}_0 \given \mbf{X}, \mbf{Y}}$ displayed as $\bigcirc$ using the Gaussian distribution in (c) and the gamma distribution in (d) as well as the underlying distribution $\p{X \mathrm{\,at\,} z}{z, \Omega}$ displayed as $\times$.}
\label{fig:F}
\end{figure}

What happens when the model used for the analysis of the data does not correspond in functional form to that of the underlying physical distribution?  Without insight into the true nature of the objects measured, there is no reason \textit{a priori} to suppose that some function chosen to resemble the data corresponds to that of the underlying physics.  Specifically, let us consider a distribution for the location $z < \bar{\phi}$ of particles of type $X$ given by \beq
f_\Omega ( z_f \given \tilde{\phi} ) = z_f^{\tilde{\phi}-1} e^{-z_f / \tilde{\phi}} / \tilde{\phi}^{\tilde{\phi}} \Gamma (\tilde{\phi}) \; ,
\eeq where $z_f \equiv \bar{\phi} - z$, and similarly for $g_\Omega ( z_g \given \tilde{\gamma} )$ using $z_g \equiv z - \bar{\gamma}$, which one may recognize as a gamma distribution with a mean of $\tilde{\phi}^2$ and a variance of $\tilde{\phi}^3$ reflected in $z$ and offset by $\bar{\phi}$.  In panels (a) and (b) of Figure~\ref{fig:F} are histograms of the locations of $X$ and $Y$ particles drawn from such a distribution with $m_\Omega = 1$, $\tilde{\phi}_\Omega = \tilde{\gamma}_\Omega = 1.5$, and $\bar{\phi}_\Omega = - \bar{\gamma}_\Omega = 4$.

Using the model of the preceding section, we get the estimates for the observable $\langle \p{X \mathrm{\,at\,} z}{z, \mbf{r}} \rangle_{\mbf{r} \given \mbf{X}, \mbf{Y}}$ and $\p{X \mathrm{\,at\,} z}{z, \mbf{r}_0 \given \mbf{X}, \mbf{Y}}$ shown in panel (c) of Figure~\ref{fig:F}.  While the observable resembles the physical distribution in the region of overlap between the particle types, the model is not designed to handle the case of finite boundaries in the location distribution.  For the unidentified particles found outside that region, this model will almost always make the wrong prediction.  One's suspicion might be aroused by noticing that the only measurements in the extreme regions are in contradiction to the model's estimate.  Of course, simply looking at the histograms reveals that the symmetric Gaussian model is not going to be the best fitting distribution for the identified particle locations.

Suppose now that somehow we gain knowledge of the functional form for $f_\Omega(z)$ and $g_\Omega(z)$, for example by learning that the particles are racquetballs hit out of a tunnel such that $X$ balls are bounced off the right wall and $Y$ balls are bounced off the left, with the measurements for the $z$ locations taken some distance away from the outlet of the tunnel.  The parameters $\bar{\phi}_\Omega$ and $\bar{\gamma}_\Omega$ are assumed to be known from the geometry of the apparatus, with the parameters $\mbf{r} = (m, \tilde{\phi}, \tilde{\gamma})$ to be determined.  The new parameter evidence is $\p{\tilde{\phi}}{\bar{\phi}, \mbf{X}} \propto \tilde{\phi}^{-1}\prod_{z \in \mbf{X}} f(z_f)$ when the prior $\p{\tilde{\phi}}{} \propto \tilde{\phi}^{-1}$ is used, and similarly for $\tilde{\gamma}$.  To remain finite as $z_f \rightarrow 0$, one requires $\tilde{\phi} \geq 1$, and the mode value $\tilde{\phi}_0$ is easily found numerically, yielding the mode estimate $\p{X \mathrm{\,at\,} z}{z, \mbf{r}_0}$, as shown in panel (d) of Figure~\ref{fig:F}.  An upper limit of $\tilde{\phi} \leq 4$ is consistent with the measurements, and the expectation of the observable $\langle \p{X \mathrm{\,at\,} z}{z, \mbf{r}} \rangle_{\mbf{r} \given \mbf{X}, \mbf{Y}}$ can be evaluated, also shown in panel (d).  Having the correct physical model, even if its parameters are undetermined, is certainly an asset when attempting to use measurements to make predictions.

The point of this section is to emphasize how important one's knowledge of the situation is to the determination of one's results.  The model one selects should be based upon as much information as is available.  When no single model presents itself as being physically correct, one must consider the alternatives, most often after having looked at the data---if the data look like an exponential decay, there is not much point in fitting a Gaussian.  The prior for its parameters likewise should draw upon that background knowledge yet remain as unbiased as possible towards the final outcome.  What Bayesian methods provide is a systematic framework for the comparison of models which all do a reasonable job of fitting the data, especially when the number or quality of measurements is low.  The transformation group principle supplements Bayes' theorem by providing a systematic framework for the evaluation of the least biased prior based upon similarity transformations of the model with respect to the data.

\section{Unknown means and deviations with measurement uncertainty}
Returning to the use of the Gaussian model for the particle locations, let us now suppose the even more realistic situation where the location measurements are themselves subject to Gaussian deviation $\sigma$, presumed to be known from calibration of the measurement device, which for convenience will be set equal to the unit for $z$ such that $\sigma \equiv 1$.  The particle locations are supposed to be drawn from independent distributions as before, whose parameters are to be determined.  The model selection ratio can be written in terms of the relative model likelihoods as \beq
\rho^{f g}_h \equiv \dfrac{\langle \p{\mbf{X}}{\bar{f}, \tilde{f}} \rangle_{\bar{f}, \tilde{f}} \langle \p{\mbf{Y}}{\bar{g}, \tilde{g}} \rangle_{\bar{g}, \tilde{g}}}{\langle \p{\mbf{X}, \mbf{Y}}{\bar{h}, \tilde{h}} \rangle_{\bar{h}, \tilde{h}}} = \dfrac{\rho_f \rho_g}{\Delta_z \Delta_{\log \sigma} \rho_h} \; ,
\eeq with the prior limits notated as before.  Focusing on the model for type $X$, each datum likelihood must now be expressed as an integral over all possible values of $X_j$ according to the resolution of the apparatus, \bes
\p{X_j}{\sigma, \bar{f}, \tilde{f}} &=& \int_{-\infty}^\infty f(z_j) \p{z_j}{\sigma, X_j} \, dz_j \\
 &=& \int_{-\infty}^\infty ( 2 \pi \tilde{f} )^{-1} \exp^{-1/2} \lbrace [ (\bar{f} - z_j) / \tilde{f}]^2 + (X_j - z_j)^2 \rbrace \, dz_j \\
 &=& [ 2 \pi (1 + \tilde{f}^2) ]^{-1/2} \exp^{-1/2} [ (\bar{f} - X_j)^2/ (1 + \tilde{f}^2) ] \; .
\ees  The parameter evidence is now given by \beq
\p{\bar{f}, \tilde{f}}{\sigma, \mbf{X}} \propto \tilde{f}^{-1} [ 2 \pi (1 + \tilde{f}^2) ]^{-J/2} \exp^{-J/2} [ (\bar{f}^2 - 2 \bar{f} \langle X_j \rangle_j +  \langle X_j^2 \rangle_j) / (1 + \tilde{f}^2) ] \; ,
\eeq retaining explicitly the normalization of the likelihood but not the prior.  The integral over $\bar{f}$ proceeds as before, \beq
\p{\tilde{f}}{\sigma, \mbf{X}} = \int_{-\infty}^\infty \p{\bar{f}, \tilde{f}}{\sigma, \mbf{X}} \, d\bar{f} \propto \tilde{f}^{-1} [ 2 \pi (1 + \tilde{f}^2) ]^{(1-J)/2} J^{-1/2} \exp^{-1/2} [ \xi_f^2 / (1 + \tilde{f}^2) ] \; ,
\eeq yielding the marginal evidence for $\tilde{f}$.  Under a change of variable $\tilde{\sigma}^2 = 1 + \tilde{f}^2$ such that $\tilde{f}^{-1} d\tilde{f} = \tilde{\sigma} (\tilde{\sigma}^2 - 1)^{-1} d\tilde{\sigma}$, the marginal evidence for $\tilde{f}$ can be rewritten as \beq
\p{\tilde{\sigma}}{\sigma, \mbf{X}} \propto (2 \pi)^{(1-J)/2} J^{-1/2} (1-\tilde{\sigma}^{-2})^{-1} \tilde{\sigma}^{-J} \exp^{-1/2} (\xi_f^2 / \tilde{\sigma}^2) \; ,
\eeq thus the remaining integral over $\tilde{\sigma}$ can be written as an infinite series, \beq
\rho_f \approx (2 \pi)^{(1-J)/2} (8 J)^{-1/2} \sum_{\alpha = 0}^\infty 2^{(J + 2 \alpha)/2} \xi_f^{1 - J - 2 \alpha} \Delta_\Gamma (\alpha,\xi_f) \; ,
\eeq where the approximation results from finite $\Delta_z$ and $\Delta_\Gamma$ now depends on $\alpha$ as well as the limits of integration $\tilde{\sigma}_{0, \infty} = (1 + \sigma^2_{0, \infty})^{1/2}$, \beq
\Delta_\Gamma (\alpha,\xi_f) \equiv \Gamma \left( \dfrac{J + 2 \alpha - 1}{2},\dfrac{\xi_f^2}{2 \tilde{\sigma}_\infty^2} \right) - \Gamma \left( \dfrac{J + 2 \alpha - 1}{2},\dfrac{\xi_f^2}{2 \tilde{\sigma}_0^2} \right) \; .
\eeq  The relative likelihoods $\rho_g$ and $\rho_h$ are evaluated similarly.

The gradient of the parameter info, \beq	
\del \q{\bar{f}, \tilde{f}}{\sigma, \mbf{X}} = \left[ \begin{array}{c} 
J ( 1 + \tilde{f}^2 )^{-1} ( \bar{f} - \langle X_j \rangle_j ) \\ 
\tilde{f}^{-1} + J ( 1 + \tilde{f}^2 )^{-1} \tilde{f} [ 1 - ( 1 + \tilde{f}^2 )^{-1} ( \bar{f}^2 - 2 \bar{f} \langle X_j \rangle_j + \langle X_j^2 \rangle_j ) ] 
\end{array} \right] \; ,
\eeq yields the same mode for the central location $\bar{f}_0 = \langle X_j \rangle_j$.  When that value is substituted into the expression for $\dsub{\tilde{f}} \q{\bar{f}, \tilde{f}}{\sigma, \mbf{X}}$, the equation for the mode of the deviation becomes \beq
0 = ( J + 1 ) \tilde{f}_0^4 + ( J + 2 - \xi ) \tilde{f}_0^2 + 1 \; ,
\eeq whose root minimizes its contribution to the parameter info, \beq
\tilde{f}_0 = \min_{\tilde{f}} \left[ \log \tilde{f} + \log ( 1 + \tilde{f}^2 )^{J/2} + \xi_f^2 / 2 ( 1 + \tilde{f}^2 ) \right] \; .
\eeq The remaining modes $\tilde{g}_0$ and $\tilde{h}_0$ are found similarly.

The uncertainty in the measurement apparatus must also be taken into account when evaluating the observable.  Given some measured location $z$ for an unidentified particle, what we know is that its actual value $z_\Omega$ is distributed around $z$ with deviation $\sigma$.  Consequently, the integration is now over not only the 5 dimensional parameter manifold but also over all possible values of $z_\Omega$, \bes
\p{X \mathrm{\,at\,} z}{z, \sigma, \mbf{X}, \mbf{Y}} &=& \int_{-\Delta_z / 2}^{\Delta_z / 2} \p{X \mathrm{\,at\,} z_\Omega}{z_\Omega, \sigma, \mbf{X}, \mbf{Y}} \p{z_\Omega}{z, \sigma} dz_\Omega \\
 &=& \int_{-\Delta_z / 2}^{\Delta_z / 2} \int_{\mbf{r}} \p{X \mathrm{\,at\,} z_\Omega}{z_\Omega, \mbf{r}} \p{\mbf{r}}{\sigma, \mbf{X}, \mbf{Y}} \p{z_\Omega}{z, \sigma} \, d\mbf{r} dz_\Omega \; ,
\ees which is evaluated numerically as before.  The restricted limits of integration are found for the independent submanifolds such that the net integration measure is approximately normalized, $\rho^\mrm{num}_\mrm{anl} \approx 1$.

\begin{figure}
\includegraphics[width=\textwidth]{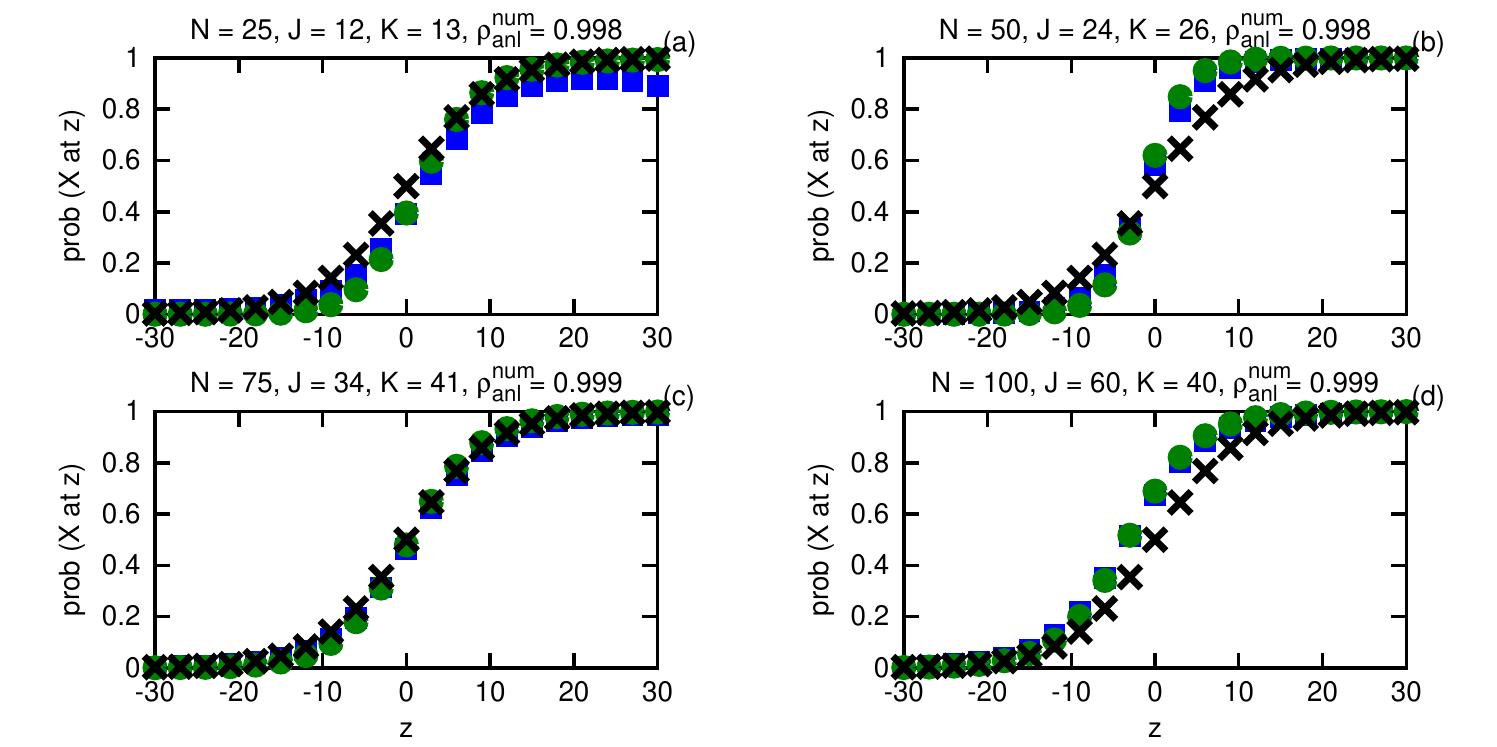}
\caption[]{Comparison of the expectation value $\langle \p{X \mathrm{\,at\,} z}{z, \sigma, \mbf{r}} \rangle_{\mbf{r} \given \sigma, \mbf{X}, \mbf{Y}}$ displayed as $\square$ to the estimate from the parameter mode $\p{X \mathrm{\,at\,} z}{z, \sigma, \mbf{r}_0 \given \sigma, \mbf{X}, \mbf{Y}}$ displayed as $\bigcirc$ as well as the underlying distribution $\p{X \mathrm{\,at\,} z}{z, \sigma, \Omega}$ displayed as $\times$ with parameter values as given in Table~\ref{tab:B}.}
\label{fig:G}
\end{figure}

In Figure~\ref{fig:G} we show the results of such an integration for $N$ particles as indicated above each panel generated using parameters found in Table~\ref{tab:B} with $\Delta_z = 60$.  We can see that, despite the additional dimension of integration, the results are quite similar to what we had before.  That should be no surprise, as the normal distribution has the feature that its mean and mode are at the same value.  If during the calibration procedure a resolution (point spread) function other than Gaussian is determined for the apparatus, it should of course be used instead.

\begin{table}
\caption{\label{tab:B} Parameters corresponding to Figure~\ref{fig:G}}
\centering
\fbox{%
\begin{tabular}{c|rrrrrrr} 
\multirow{2}{*}{panel} & $\log_{10} \rho^{f g}_h$ & $N$ & $m_\Omega$ & $\bar{f}_\Omega$ & $\bar{g}_\Omega$ &  $\tilde{f}_\Omega$ & $\tilde{g}_\Omega$ \\
 & $J$ &  $K$ & $m_0$ & $f_0$ & $g_0$ & $\tilde{f}_0$ & $\tilde{g}_0$ \\ \hline
\multirow{2}{*}{a} &  1.99 &    25 &     1 &    10 &   -10 &    10 &    10 \\
 &    12 &    13 & 0.923 &  10.7 & -7.41 &  7.38 &  9.25 \\ \hline
\multirow{2}{*}{b} &  11.1 &    50 &     1 &    10 &   -10 &    10 &    10 \\
 &    24 &    26 &     1 &  11.3 & -14.1 &  7.56 &  8.02 \\ \hline
\multirow{2}{*}{c} &  8.98 &    75 &     1 &    10 &   -10 &    10 &    10 \\
 &    34 &    41 &  1.14 &  8.98 & -9.45 &  8.64 &  9.08 \\ \hline
\multirow{2}{*}{d} &  14.9 &   100 &     1 &    10 &   -10 &    10 &    10 \\
 &    60 &    40 & 0.639 &  9.59 & -12.5 &  9.55 &  9.52 \\ 
\end{tabular}}
\end{table}

\section{Reliability of the estimate}
So far we have danced around the topic of determining the reliability of the estimate strictly from the measurements at hand.  In all the previous figures we have displayed the underlying physical distribution as a means of establishing that the method does indeed approach the ``true'' value as the number of measurements increases.  However, in the real world, that knowledge is beyond our ken; we must make do with what the data have to say for themselves.  That information is encoded in the evidence density for the model parameters given the measurements and the resolution of the apparatus.

When sufficient data exist that the significant evidence is restricted to some tiny region around a mode $\mbf{r}_0$ that barely moves as more data is collected, then we may as well call the prediction from the mode our single best estimate of the underlying distribution, $\p{X \mathrm{\,at\,} z}{z, \sigma, \mbf{r}_0} \approx \p{X \mathrm{\,at\,} z}{z, \sigma, \Omega}$.  In that case, the chance of making the correct prediction $P \in \{X,Y\}$ for some new datum $O \in \{X,Y\}$ is given by a simple truth table.  Using the notation $x_\Omega \equiv \p{X \mathrm{\,at\,} z}{z, \sigma, \Omega}$, we have \beq \label{eqn:xomcorr}
p ( O = P \given z, \sigma, \mbf{r}_0 ) \approx \mrm{Tr} \left[ \begin{array}{cc} x_\Omega^2 & x_\Omega (1-x_\Omega) \\ x_\Omega (1-x_\Omega) & (1-x_\Omega)^2 \end{array} \right] = x_\Omega^2 + (1-x_\Omega)^2 \; ,
\eeq which indicates that even knowledge of the physical distribution does not guarantee a certain prediction for the particle type of the new datum; to be absolutely certain of the new particle's type for any location, one must measure its classification.  The error rate is greatest at the location where the particles appear with equal likelihood, as the chance of a successful prediction there is 1/2.  When the mode of the model parameters is extremely well determined by the data, the error rate lies between 0 and 50\% according to the value of $\p{X \mathrm{\,at\,} z}{z, \sigma, \mbf{r}_0}$.

While the prediction from the mode is the most likely contribution, the prediction from the mean of the observable is what encodes our best inference about its value into a single number.  If we want to know more about the observable, we have to do more work.  As mentioned earlier, to describe the distribution around the expected value of the observable at some location, we need an additional parameter for its width and some function for its shape.  The natural distribution for unit normalized positive quantities is the beta distribution $\p{x}{a, b} = x^{a-1} (1-x)^{b-1} / \beta(a,b)$ that we have encountered in various guises already.  Here, the problem is to determine its parameters $a_z$ and $b_z$ given empirical knowledge of the distribution of $x_z \equiv \p{X \mathrm{\,at\,} z}{z, \sigma, \mbf{X}, \mbf{Y}}$.

One can easily show that the maximum likelihood estimate of the parameters is given by the solution of the system of equations \bea
\Lambda_1 (a_z) - \Lambda_1 (a_z+b_z) &=& \langle \log x_z \rangle \; , \label{eqn:foraz} \\
\Lambda_1 (b_z) - \Lambda_1 (a_z+b_z) &=& \langle \log (1-x_z) \rangle \; , \label{eqn:forbz}
\eea using the notation $\Lambda_k (r) \equiv (\dsub{r})^k \log \Gamma (r)$ for the polygamma functions with integer order $k$ and real argument $r$,  which selects the parameters for the beta distribution whose values of $\langle \log x_z \rangle$ and $\langle \log (1-x_z) \rangle$ equal those estimated from the empirical measurements.  We set aside (for this article) the question of whether a non-uniform prior $\p{a, b}{}$ should be incorporated here on the grounds that a lot of effort will be put into converging the integrals such that the prior should have little effect.

The task now is to evaluate those expectation values conditioned on the given set of data.  Instead of a single observable at each displayed location, there are now two to calculate, the first given by \beq
\langle \log x_z \rangle = - \int_{-\Delta_z / 2}^{\Delta_z / 2} \int_{\mbf{r}} \log [ 1 + m \zeta ( z_\Omega ) ] \p{\mbf{r}}{\sigma, \mbf{X}, \mbf{Y}} \p{z_\Omega}{z, \sigma} \, d\mbf{r} dz_\Omega \; ,
\eeq and the second given by \beq
\langle \log ( 1 - x_z ) \rangle = \langle \log x_z \rangle + \int_{-\Delta_z / 2}^{\Delta_z / 2} \int_{\mbf{r}} \log [ m \zeta ( z_\Omega ) ] \p{\mbf{r}}{\sigma, \mbf{X}, \mbf{Y}} \p{z_\Omega}{z, \sigma} d\mbf{r} dz_\Omega \; .
\eeq   From those two estimates one finds the corresponding $a_z$ and $b_z$ according to Equations~(\ref{eqn:foraz}) and (\ref{eqn:forbz}) above.  The integral over $z_\Omega$ in the second term can be expressed analytically when taken over infinite limits, \beq
\int_{-\infty}^\infty \log [ m \zeta ( z_\Omega ) ] \p{z_\Omega}{z, \sigma} dz_\Omega = \log \left( \dfrac{m \tilde{f }}{\tilde{g}} \right) + \dfrac{( z - \bar{f} )^2 + 1}{2 \tilde{f}^2} - \dfrac{( z - \bar{g} )^2 + 1}{2 \tilde{g}^2} \; ,
\eeq thereby reducing the amount of effort required for its evaluation.  Having found $a_z$ and $b_z$, the mean of the observable $x_z$ may be determined according to $\langle x_z \rangle = a_z / (a_z+b_z)$, and its mode is given by $x_{z, 0} = (a_z-1) / (a_z+b_z-2)$ when $a_z, b_z \geq 1$ else may be found at 0 or 1.

\begin{figure}
\includegraphics[width=\textwidth]{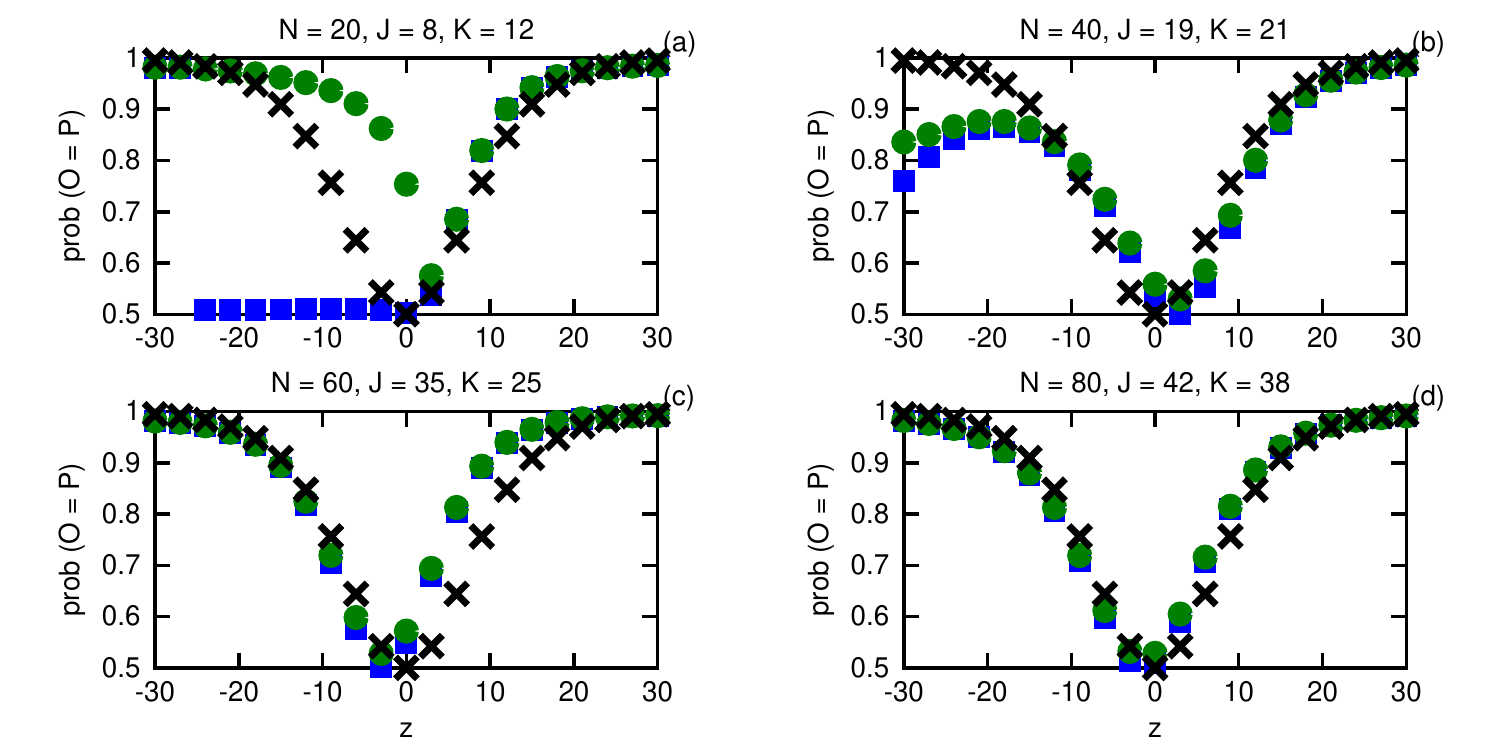}
\caption[]{Comparison of the empirical success rates given by Equation~(\ref{eqn:xzcorr}) displayed as $\square$ and by Equation~(\ref{eqn:azbzcorr}) displayed as $\bigcirc$ to the optimal success rate displayed as $\times$ with parameter values given in Table~\ref{tab:C}.}
\label{fig:H}
\end{figure}

\begin{table}
\caption{\label{tab:C} Parameters corresponding to Figure~\ref{fig:H}}
\centering
\fbox{%
\begin{tabular}{c|rrrrrrr} 
\multirow{2}{*}{panel} & $\log_{10} \rho^{f g}_h$ & $N$ & $m_\Omega$ & $\bar{f}_\Omega$ & $\bar{g}_\Omega$ &  $\tilde{f}_\Omega$ & $\tilde{g}_\Omega$ \\
 & $J$ &  $K$ & $m_0$ & $f_0$ & $g_0$ & $\tilde{f}_0$ & $\tilde{g}_0$ \\ \hline
\multirow{2}{*}{a} &  2.72 &    20 &     1 &    10 &   -10 &    10 &    10 \\       
                   &     8 &    12 &  1.22 &  10.5 & -13.6 &  6.24 &  9.52 \\ \hline
\multirow{2}{*}{b} &   3.7 &    40 &     1 &    10 &   -10 &    10 &    10 \\       
                   &    19 &    21 &     1 &  10.9 & -6.62 &  9.95 &  7.73 \\ \hline
\multirow{2}{*}{c} &   8.6 &    60 &     1 &    10 &   -10 &    10 &    10 \\       
                   &    35 &    25 & 0.667 &  9.21 & -11.6 &  8.85 &  8.41 \\ \hline
\multirow{2}{*}{d} &  11.5 &    80 &     1 &    10 &   -10 &    10 &    10 \\       
                   &    42 &    38 &  0.86 &  10.6 & -12.3 &  10.2 &  9.86 \\       
\end{tabular}}
\end{table}

When using the empirical distribution to make predictions, the chance of a successful prediction is itself given by an expectation value.  The closest thing we have to knowledge of $x_\Omega$ is what we know about the distribution of $x_z$.  If one uses $\langle x_z \rangle$ as the basis of prediction, by comparing it to a uniformly drawn random deviate $u$, then the chance of a successful prediction is given by \bes \label{eqn:xzcorr}
\langle p ( O = P \given z, \sigma, \langle x_z \rangle ) \rangle &=& \langle x_z \langle x_z \rangle \rangle + \langle (1 - x_z) \langle 1 - x_z \rangle \rangle =  \langle x_z \rangle^2 + ( 1 - \langle x_z \rangle )^2 \\
 &=& \dfrac{a_z^2 + b_z^2}{(a_z + b_z)^2} \; ,
\ees which amounts to inserting the expectation value of $x_z$ into the expression for the optimal success rate, Equation~(\ref{eqn:xomcorr}).  We can, however, do better than that.  If we compare $u$ instead to a value $x$ drawn from the distribution $\p{x}{a, b}$ at $z$, the chance of a successful prediction is then given by \bes \label{eqn:azbzcorr}
\langle p ( O = P \given z, \sigma, a_z, b_z ) \rangle &=& \langle x_z^2 + ( 1 - x_z)^2 \rangle = \langle x_z^2 \rangle + \langle ( 1 - x_z)^2 \rangle \\
 &=& \dfrac{a_z^2 + b_z^2 + a_z + b_z}{(a_z+b_z) (a_z + b_z + 1)} \; ,
\ees which is greater than or equal to the success rate based on $\langle x_z \rangle$ for all $a_z, b_z > 0$.  As promised, the Bayesian methodology has delivered an estimate of the observable used for predicting the type of unclassified particles as well as an estimate of its error rate based entirely upon the data at hand.  In Figure~\ref{fig:H} we compare the empirical success rates of Equations~(\ref{eqn:xzcorr}) and (\ref{eqn:azbzcorr}) with the optimal success rate of Equation~(\ref{eqn:xomcorr}), for $N$ particles as indicated above each panel generated using parameters found in Table~\ref{tab:C}.

\section{Discussion}
The purpose of the preceding exercises is to demonstrate how Bayes' theorem and the principle of indifference are used to extract meaningful information from a set of data.  When the observable representing the desired knowledge is not one of the parameters of the model, then the single best inference of its value is given by its expectation over the parameter manifold weighted by the evidence measure.  While the estimate from the parameter mode is the single most likely contribution, it is the mean of the observable which takes into account how much evidence there is in the data for other possibilities.  When the transformation group principle is used to find the invariant measure for the parameter manifold, the results of a Bayesian analysis are the least biased possible, as each datum updates the evidence density according to its contribution to the available information which starts with the statement of the geometric properties of the model with respect to the data.

One feature of Bayesian data analysis as expressed in the language of conditional probabilities is that it forces one to specify the state of background knowledge upon which any estimate is based.  Whenever real data are discussed, such statements necessarily include remarks on the nature of the measurement apparatus.  That apparatus can be in the form of either a physical device such as a voltmeter or an abstract device such as a survey.  While the parameter $z$ has been given the interpretation of a spatial coordinate, it represents any observable that can be expressed in terms of a location-type parameter with uniform measure over an arbitrarily large domain.  That parameter can of course be generalized to a data vector of any dimensionality according to the complexity of the situation at hand.

When one is not certain that the background knowledge includes specification of the physically correct functional form for the constituent distributions, there is a formal procedure for comparing the relative likelihood of competing models.  That comparison is expressed not by the ratio of the peak likelihoods given by each model's parameter mode but rather by the ratio of the expected likelihoods given by the integral of the evidence over each model's parameter manifold.  The distinction is important, as it is the latter which takes into account the principle of efficiency through the Occam factor.  The most useful form of the parameter evidence density is that which drops the normalization of the prior yet retains any constant factors in the likelihood, as its integral appears as both the relative likelihood of the model and the normalization constant for taking expectation values of observables.  Care must be taken when dropping constants from an evaluation (whether numerical or analytic), as it is easy to make a mistake regarding the appropriate normalization for the given task.

When not all of the prior normalization factors cancel out of the model comparison ratio, then they must be accounted for explicitly.  Quite often the transformation group prior will yield an infinite normalization when allowed to extend over an infinite domain, which is no less a problem for a maximum likelihood analysis that assigns any and all parameters a uniform prior.  In that case, one is forced to consider more fully the nature of the apparatus as well as of the objects to be measured in order to determine sensible boundaries for one's prior state of knowledge.  One generally hopes that finite boundaries do not truncate significantly the evidence density, but there can be times when physical constraints (such as positivity) dictate that what would otherwise be the mode lies outside the allowable parameter domain.

The process of inductive reasoning, through which one expresses one's degree of belief in the value of some observable, is not restricted to the simple distributions considered here but extends naturally to much more complicated situations.  By going through the development of the methodology for increasingly less restrictive states of prior knowledge, we hope that readers have gained some insight into how to apply the method to problems they encounter.  While almost everyone is comfortable with the process of finding the parameter mode, and most with taking expectation values according to the evidence density, the process of model selection often seems mysterious until one realizes that the relevant factors are just the relative expected likelihoods and any leftover prior normalizations.

\section{Conclusion}
In this article we have explored the transformation group approach to the problem of predicting the type of some unclassified particle given its location and a list of locations for particles of identified classification.  The process of inductive reasoning is used to relate these quantities of interest, which incorporates both Bayes' theorem and the principle of indifference to produce the least biased estimate from the data at hand.  The expectation value of the observable is compared to that derived from the parameter mode, and they converge on the underlying distribution when sufficient data exists and the model function is known to be physically correct.  When competing models must be compared, the procedure of evaluating the evidence ratio in terms of the relative likelihoods and prior normalizations has been explicitly determined.

The prediction from the Bayesian method is found to be superior to those from non-parametric algorithms, such as nearest neighbor or kernel density estimation, especially when not much data exists for analysis.  The reason is because conditional probability theory makes full use of all the available information rather than discarding the contribution from some datum on the grounds that it is far away from the desired location or smearing that information out according to some linear operator.  The essential difference between inductive and deductive methods is that the former inverts the model to predict the data whereas the latter inverts the data to predict the model---see Ref.~\cite{johnson:70233} for an explicit example regarding sums of exponential functions and Ref.~\cite{rwj:astro03} as regards the Fourier transform.  Having invested in the presumably expensive classification of the particles in the training set, it makes sense that one should choose the most reliable method for analyzing that data.  It is only those of us who generate data cheaply that worry about the expense of computing an integral.

A major advantage of the Bayesian method is that it is extensible.  As new effects are brought into play, their parameters are included in the analysis in a straightforward manner.  The language of conditional probability theory is sufficiently general that it can handle whatever state of background knowledge one specifies, including those which admit that competing models are available as well as those which admit that the measurements are themselves subject to error.  ``Reasoning in the face of uncertainty'' is an apt description of inferential logic and is a problem often faced in the imperfect real world.  The use of Bayes' theorem requires one to specify the conditions upon which any and all probabilities are based, which seems like a lot of work at the outset but leads to a robust analysis upon completion.  It also recognizes the difference between the most likely value of some observable and its expected value, which is important to consider when not many measurements are available.  By working through the tasks of model selection, parameter estimation, and observable prediction explicitly for the case of these simple distributions, we hope the reader has learned how to apply the transformation group method to more complicated cases of data analysis.

\nolinenumbers


\bibliographystyle{cJAS}       


\label{lastpage}

\end{document}